\documentclass[bibyear]{aa} 
\usepackage{graphicx}
\usepackage{txfonts}
\usepackage{here} 
\def\f#1   {Fig.~\ref{#1}}
\def\s#1   {Sec.~\ref{#1}}
\def\tab#1   {Table~\ref{#1}}
\def\t#1   {Tab.~\ref{#1}}

\def\comm#1   {{\tt (COMMENT: #1) }}
\def\kms{~km~s$^{\mathrm{-1}}$}

\def\i                {{\em i}}

\def\lsol              {$\mathrm{L}_{\odot}$}

\def\msol              {$\mathrm{M}_{\odot}$}

\def\msolyr              {$\mathrm{M}_{\odot}\, \mathrm{yr}^{-1}$}

\def\smo               {Smol\v{c}i\'{c}}

\begin{document} 

\title{Physical properties of $z>4$ submillimeter galaxies in the COSMOS field}

   \author{
   V.~Smol\v{c}i\'{c}\inst{1}, 
   A.~Karim\inst{2}, 
   O.~Miettinen\inst{1}, 
   M.~Novak\inst{1}, 
   B.~Magnelli\inst{2}, 
   D.~A.~Riechers\inst{3}, 
   E.~Schinnerer\inst{4}, 
   P.~Capak\inst{5,6}, 
   M.~Bondi\inst{7}, 
   P.~Ciliegi\inst{8}, 
   M.~Aravena\inst{9}, 
   F.~Bertoldi\inst{2}, 
   S.~Bourke\inst{5}, 
   J.~Banfield\inst{10,11},
   C.~L.~Carilli\inst{12},
   F.~Civano\inst{13,14}, 
   O.~Ilbert\inst{15}, 
   H.~T. Intema\inst{12},
   O.~Le~F\`{e}vre\inst{15}, 
   A.~Finoguenov\inst{16}, 
   G.~Hallinan\inst{5}, 
   H.-R.~Kl\"{o}ckner\inst{17},
   C.~Laigle\inst{18,19}, 
   D.~Masters\inst{20}, 
   H.~J.~McCracken\inst{18,19}, 
   K.~Mooley\inst{5}, 
   E.~Murphy\inst{5}, 
   F.~Navarette\inst{2}, 
   M.~Salvato\inst{21}, 
   M.~Sargent\inst{22},
   K.~Sheth\inst{23},
   S.~Toft\inst{24},
   G.~Zamorani\inst{8}
    }

   \institute{
  Department of Physics, University of Zagreb, Bijeni\v{c}ka cesta 32, HR-10000 Zagreb, Croatia 
\and 
  Argelander Institute for Astronomy, Auf dem H\"{u}gel 71, Bonn, 53121, Germany 
\and
Department of Astronomy, Cornell University, 220 Space Sciences Building, Ithaca, NY 14853, USA 
\and 
Max Planck Institut f\"{u}r Astronomie, K\"{o}nigstuhl 17, 69117 Heidelberg, Germany 
\and
Department of Astronomy, California Institute of Technology, MC 249-17, 1200 East California Blvd, Pasadena, CA 91125, USA
\and 
Spitzer Science Center, 314-6 Caltech, Pasadena, CA 91125, USA 
\and 
  INAF - Istituto di Radioastronomia, via Gobetti 101, I-40129 Bologna, Italy 
\and 
  INAF - Osservatorio Astronomico di Bologna, via Ranzani 1, 40127 Bologna, Italy 
\and 
N\'ucleo de Astronom\'{\i}a, Facultad de Ingenier\'{\i}a, Universidad Diego Portales, Av. Ej\'ercito 441, Santiago, Chile
\and
CSIRO Australia Telescope National Facility, PO Box 76, Epping, NSW, 1710, Australia
\and
Research School of Astronomy and Astrophysics, Australian National University, Weston Creek, ACT, 2611, Australia
\and
 National Radio Astronomy Observatory, P.O. Box 0, Socorro, NM 87801, USA
\and
Yale Center for Astronomy and Astrophysics, Physics Department, Yale University, PO Box 208120, New Haven, CT 06520-8120, USA 
\and 
Harvard-Smithsonian Center for Astrophysics, 60 Garden Street, Cambridge, MA 02138, USA 
\and 
  Aix Marseille Universit\'{e}, CNRS, LAM (Laboratoire d'Astrophysique de Marseille) UMR 7326, 13388, Marseille, France 
\and 
  Department of Physics, University of Helsinki, P.O. Box 64, FI-00014, Helsinki, Finland 
\and 
Max-Planck-Institut f\"{u}r Radioastronomie, Auf dem H\"{u}gel 69 D-53121 Bonn, Germany 
\and 
Sorbonne Universit\'{e}s, UPMC Univ Paris 06, UMR 7095, Institut d'Astrophysique de Paris, F-75005, Paris, France 
\and 
  Institut d'Astrophysique de Paris, UMR 7095 CNRS, Universit\'{e} Pierre et Marie Curie, 98 bis Boulevard Arago, F-75014 Paris, France 
\and 
Department of Physics and Astronomy, University of California, Riverside, 
CA 92521
\and 
Max-Planck-Institut f\"{u}r Extraterrestrische Physik, Postfach 1312, D-85741, Garching bei M\"{u}nchen, Germany 
\and 
 Astronomy Centre, Department of Physics and Astronomy, University of Sussex, 
Falmer, Brighton BN1 9QH, UK
\and 
National Radio Astronomy Observatory, 520 Edgemont Road, Charlottesville, VA 22903, USA
\and 
  Dark Cosmology Centre, Niels Bohr Institute, University of Copenhagen, Juliane Mariesvej 30, DK-2100 Copenhagen, Denmark
}

   \date{Received ; accepted}

\authorrunning{V.\ Smol\v{c}i\'{c} et al.}
\titlerunning{$z>4$ COSMOS SMGs}

\abstract{
We investigate the physical properties 
of a sample of six submillimeter galaxies (SMGs) in the COSMOS field, 
spectroscopically confirmed to lie at redshifts $z>4$. While the redshifts for 
four of these SMGs have been previously known, we present here two newly 
discovered $z_\mathrm{spec}>4$ SMGs. For our analysis we employ the rich (X-ray to radio) COSMOS 
multi-wavelength datasets. 
In particular, we use new Giant Meterwave Radio Telescope (GMRT) 325 MHz and 
3 GHz Jansky Very Large Array (VLA) data to probe the rest-frame 1.4 GHz 
emission at $z=4$, and to estimate the sizes of the star-forming regions of 
these sources, respectively.
We find that only one SMG is clearly resolved at a resolution of 
$0\farcs6\times0\farcs7$ at 3~GHz, two may be marginally resolved, while the 
remaining three SMGs are unresolved 
at this resolution. Combining this with sizes from high-resolution (sub-)mm observations available in the literature for AzTEC~1 and AzTEC~3 we infer a median radio-emitting size for our $z>4$ 
SMGs of $( 0\farcs63 \pm 0\farcs12 ) \times (0\farcs35 \pm 0\farcs05 )$ or 
$4.1 \times 2.3$ kpc$^2$ (major~$\times$~minor axis;  assuming $z=4.5$) or lower if we take the 
two marginally resolved SMGs as unresolved. This is consistent with the sizes 
of star-formation regions in lower-redshift SMGs, and local normal galaxies, 
yet higher than the sizes of star-formation regions of local ULIRGs.
Our SMG sample consists of a fair mix of compact and more clumpy systems with 
multiple, perhaps merging, components. With an average formation time of 
$\sim280$ Myr, as derived through modeling of the UV-infrared (IR) spectral 
energy distributions (SEDs), the studied SMGs are young systems. 
The average stellar mass, dust temperature, and IR luminosity we derive are 
$M_{\star}\sim1.4\times 10^{11}$ M$_{\sun}$, $T_{\rm dust}\sim43$ K, and $L_{\rm IR}\sim 1.3\times10^{13}$ 
L$_{\sun}$, respectively. The average $L_{\rm IR}$ is up to an order of 
magnitude higher than for SMGs at lower redshifts.   Our SMGs follow the correlation between dust temperature and IR luminosity as 
derived for \textit{Herschel}-selected $0.1<z<2$ galaxies. We study the 
IR-radio correlation for our sources and find a deviation from that derived 
for $z<3$ ULIRGs ($\langle q_\mathrm{IR} \rangle=1.95\pm0.26$ for our sample, 
compared to $q\approx2.6$ for IR luminous galaxies at $z<2$). In summary, we find that the physical properties derived 
for our $z>4$ SMGs put them at the high end of the $L_{\rm IR}$--$T_{\rm dust}$ 
distribution of SMGs, and that our SMGs form a morphologically heterogeneous sample. Thus, further in-depth analyses of large, 
statistical samples of high-redshift SMGs are needed to fully understand their 
role in galaxy formation and evolution.
}


\maketitle

%

\section{Introduction}
\label{sec:intro}

With extreme star-formation rates (SFRs) of $\sim10^3$~M$_{\sun}$~yr$^{-1}$ (e.g. \cite{blain2002}) (sub-)millimeter-selected galaxies (SMGs) trace a phase of the most intense stellar mass build-up in cosmic history. 
Submillimeter galaxies at $z\gtrsim4$ are now being identified at high pace (\cite{capak2008}, 2011; \cite{daddi2009a},b; \cite{coppin2009}; \cite{knudsen2010}; \cite{riechers2010}; \cite{smolcic2011}; \cite{carilli2011}; \cite{cox2011}; \cite{combes2012}; \cite{walter2012}; \cite{riechers2013}) although previously unexpected to exist in large quantities (e.g. \cite{chapman2005}).  It has been  demonstrated that a longer-wavelength selection of the population introduces a bias toward higher redshifts (as for a fixed temperature the SED dust peak occurs at longer observed wavelength bands for higher redshifts; \cite{smolcic2012}; \cite{weiss2013}; \cite{casey2013}; \cite{dowell2014}; \cite{simpson2014}; \cite{swinbank2014}; \cite{zavala2014}). Furthermore, the recent emergence of statistical samples of SMGs with accurate counterparts and redshifts has allowed us to quantitatively put SMGs in the context of massive galaxy formation (e.g. \cite{toft2014}). Based on various arguments, such as redshift distributions, surface density, maximal starburst, stellar mass, and effective radius considerations, Toft et al.\ (2014) have presented a  link between $z>3$ SMGs and the population of compact, quiescent galaxies at $z\sim2$, implying that the first are the progenitors of the latter population. 

Studies in recent years have made tremendous progress in disentangling the single-dish selected SMG population. Observational studies of large samples at improved spatial resolution (e.g., \cite{younger2007}, 2009; \cite{smolcic2012}; \cite{hodge2013}) as well as in-depth studies of carefully-selected examples (e.g., \cite{tacconi2006}, 2008; \cite{engel2010}; \cite{riechers2011}, 2013, 2014; \cite{ivison2011}; \cite{hodge2012}; \cite{debreuck2014}) have revealed that these enigmatic systems appear to entail a mix of merger-induced starbursts, physically unrelated pairs or multiples of galaxies blended in the large single-dish beams (FWHM$\sim 10\arcsec-30\arcsec$), or even extended, isolated disk galaxies (see  \cite{hayward2012}). However, the fractional contribution of these different populations, as well as their dependence on submillimeter flux, redshift, or the wavelength of selection remain subject to further study. Moreover, it remains unclear to what degree the most distant examples identified at $z>4$ are driven by different mechanisms than the more typical $z\sim2$ SMGs, and thus, if they may partially represent a different population (e.g., \cite{daddi2009a}; \cite{capak2011}).
This highlights the importance of defining well-selected samples of SMGs at a range of submillimeter fluxes and at different redshifts, in particular at the highest redshifts ($z>4$), where serendipity played an important role in the identification of a significant fraction of the few systems known (e.g., \cite{daddi2009a},b).

Micha{\l}owski et al. (2010) modeled the UV to radio spectral energy distributions (SEDs) of six spectroscopically confirmed SMGs at $z>4$. They found SFRs, stellar and dust masses, extinction and gas-to-dust mass ratios consistent with those for SMGs at $1.7<z<3.6$. They also found that the infrared (IR)-to-radio luminosity ratios for SMGs do not change on average up to $z\sim5$, yet that they are on average a factor of $\sim2.1$ lower than those inferred for local galaxies (see also, e.g., \cite{kovacs2006}; \cite{murphy2009}). Barger et al. (2012), based on their complete sample of SCUBA 850 $\mu$m sources in GOODS-N brighter than 3~mJy, found that their SMGs obey the local FIR-radio correlation.
Recently, Huang et al. (2014) performed an analysis of the SEDs in the IR (using \textit{Spitzer} and \textit{Herschel} data) for seven bright, publicly available SMGs, spectroscopically confirmed to lie at $z>4$ and drawn from various fields.\footnote{Note that five SMGs in their sample overlap with the sample analyzed by Micha{\l}owski et al. (2010), and three with the sample presented here (AzTEC~1, AzTEC~3, and J1000+0234).} They find a far-IR-radio luminosity ratio lower than that found locally. They find that their SMGs at $z>4$ are hotter and more luminous in the IR, but otherwise similar to SMGs at $z\sim2$.

Here we present an analysis of the physical properties of a sample of six $4.3 < z < 5.5$ SMGs with spectroscopic redshifts drawn from the Cosmic Evolution Survey (COSMOS) field (AzTEC1, AzTEC3, AzTEC/C159, AK03, Vd-17871, and J1000+0234). This represents the currently largest sample of $z>4$ SMGs, drawn from a single field with rich, and uniform multi-wavelength coverage. The SMGs were identified via dedicated follow-up observations  of high-redshift SMG candidates in the COSMOS field using (sub)mm bolometers (AzTEC, MAMBO) and interferometers (PdBI, SMA, CARMA), as well as optical/mm-spectroscopy [Keck~II/DEep Imaging Multi-Object Spectrograph (DEIMOS), PdBI; \cite{capak2008}, 2011; \cite{schinnerer2008}; \cite{riechers2010}; \cite{smolcic2011}; this work]. While the sources AzTEC1, AzTEC3, J1000+0234, and Vd-17871 have already been analyzed and discussed in detail elsewhere (\cite{capak2008}, 2011; \cite{schinnerer2008}; \cite{riechers2010}; \cite{smolcic2011}; \cite{riechers2014}; A.~Karim et al., in prep.), here we present two more COSMOS SMGs at $z>4$ (AzTEC/C159 and AK03). The basic properties of the sources are listed in Table~\ref{table:sample}, their multi-wavelength stamps are shown in Fig.~\ref{fig:stamps}, and notes on individual sources are given in Section~\ref{sec:sample}. 
In Section~\ref{sec:data} \ we describe the ancillary data available. 
In Section~\ref{sec:results} \ we analyze SEDs of our $z>4$ SMGs. We discuss our results in Section~\ref{sec:discussion}, and summarize them in Section~\ref{sec:summary}. For the Hubble constant, matter density, and dark energy density we adopt the values $H_0=70$~\kms ~Mpc$^{-1}$, $\Omega_{\rm M}=0.3$, and $\Omega_\Lambda=0.7$, respectively, and use 
a Chabrier (2003) initial mass function (IMF).

\begin{figure*}[H]
\includegraphics[bb=14 14 500 748, scale=0.87]{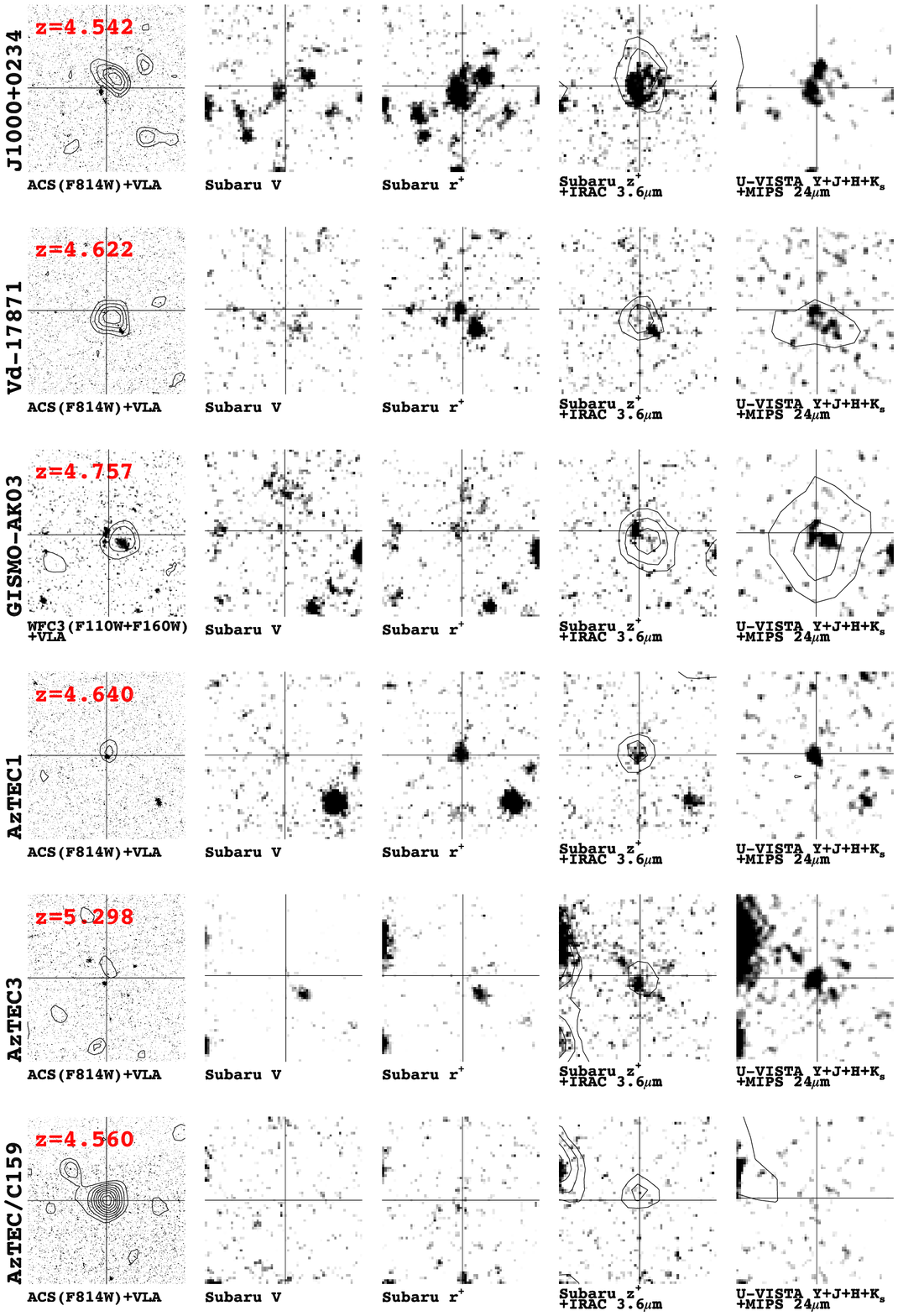}
\centering
  \caption{\footnotesize 
  xxx
 \label{dummy}
}
\end{figure*}

\begin{figure*}
\setcounter{figure}{0}
\includegraphics[bb=14 14 500 748, scale=0.87]{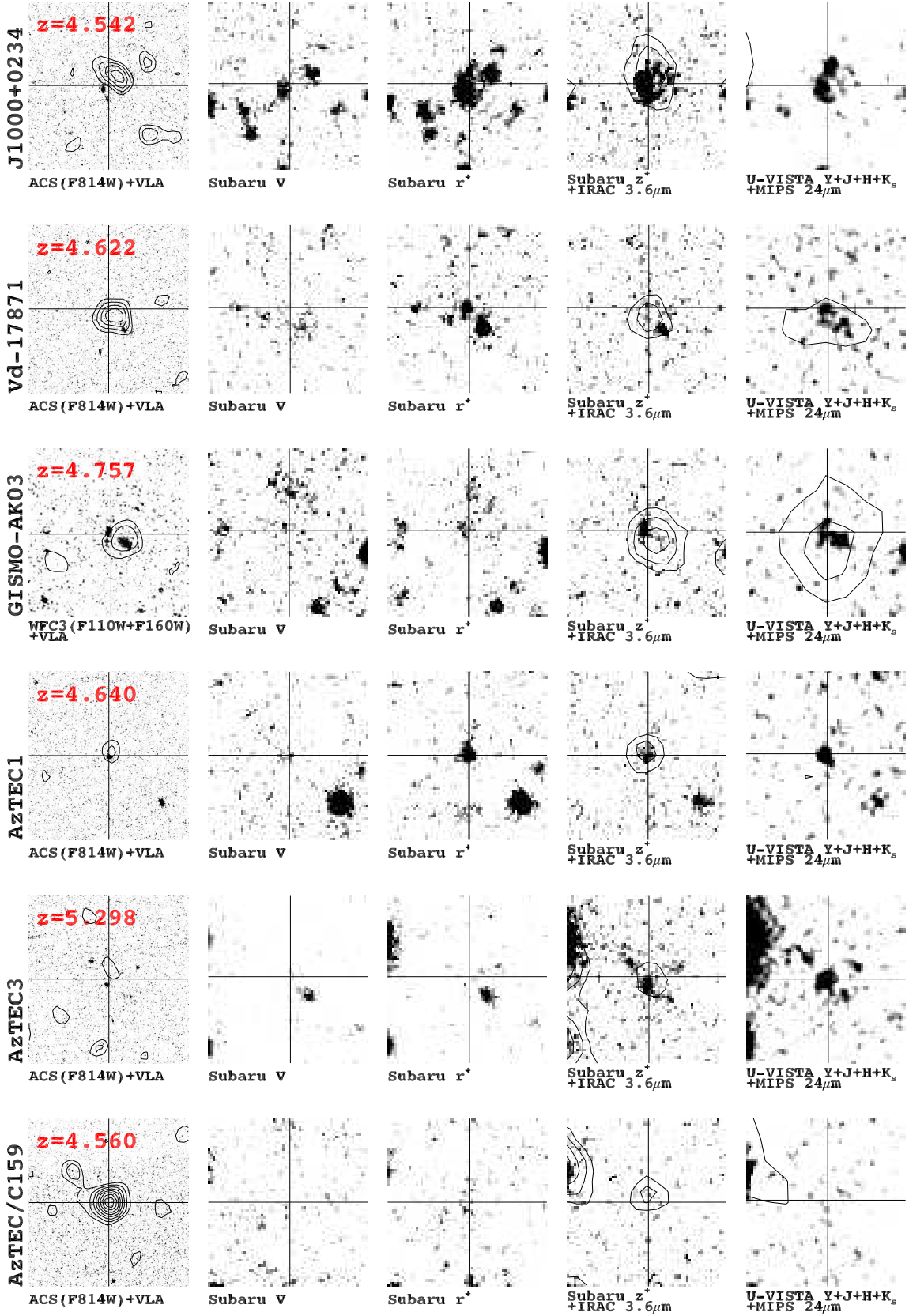}
\centering
  \caption{\footnotesize $10\arcsec \times 10\arcsec$ cut-outs of our
  $z > 4$ COSMOS SMGs. Shown are the \textit{HST}/ACS $I$-band
    (rest-frame UV) or the stacked \textit{HST}/WFC3 bands from CANDELS (PI: S. Faber; if available) overlaid with VLA 1.4 GHz radio continuum contours (levels: $2^i\sigma$, where $i=1,\,2,\,3,\ldots$, and $\sigma$ is the local rms), the Subaru $V$, $r^+$ and $z^+$ broad band filters as well as a stack of the near-IR filters of the UltraVISTA survey ($Y,\,J,\,H,\,K_{\rm s}$). \textit{Spitzer}/IRAC 3.6 $\mu$m and MIPS 24 $\mu$m contours (starting at 3$\sigma$, in steps of 3$\sigma$) are overlaid onto each $z^+$ and UltraVISTA image, respectively. Spectroscopic redshifts based on Ly$\alpha$ diagnostics from Keck/DEIMOS observations  targeting sources indicated by the large black cross are given in the top left corner of the leftmost panels. 
 \label{fig:stamps}
}
\end{figure*}

\section{Sample: Notes on individual sources}
\label{sec:sample}

\begin{itemize}
\item[$\bullet$] {\em AzTEC1} was initially detected in the JCMT/AzTEC 1.1~mm survey reaching an angular resolution of $18\arcsec$ over a 0.3~deg$^2$. area 
in the COSMOS field  (\cite{scott2008}).\footnote{All the sources in the paper by Scott et al. (2008) were identified from the central 0.15 deg$^2$ area with a uniform sensitivity of $\sim1.3$ mJy~beam$^{-1}$.} It has a statistically corrected deboosted 1.1~mm flux density of $9.3\pm1.3$~mJy (the deboosting factor is $\sim1.2$). The multi-wavelength counterpart to the millimeter source has been identified via higher resolution ($2\arcsec$) follow-up at 890~$\mu$m with the SMA (\cite{younger2007}). Combining the COSMOS multi-wavelength photometry with the Keck~II DEIMOS spectrum obtained for AzTEC1, Smol\v{c}i\'{c} et al. (2011) inferred a redshift of $z=4.64^{+0.06}_{-0.08}$ for the source, with a possible second redshift solution at $z\sim4.4$. The source's redshift has recently been 
spectroscopically determined based on detection of CO(4-3) and CO(5-4) transitions with the Large Millimeter Telescope (LMT), and then confirmed to be at $z=4.3415$ based on a $[\ion{C}{II}]$ line detection with the SMA (Yun et al., in prep). In an SMA follow-up study at $0\farcs2$ resolution the source is unresolved (\cite{younger2008}) implying an upper limit to the diameter of its star formation region of $\lesssim2~$kpc.
\item[$\bullet$] {\em AzTEC3} was initially detected in the same survey by Scott et al. (2008) as AzTEC1, and its counterparts at other wavelengths were identified by Younger et al. (2007). 
The deboosted 1.1~mm flux density is $5.9\pm1.3$~mJy (where the deboosting factor is $\sim1.3$).
Based on Keck~II DEIMOS spectra Capak et al.\ (2011) have shown that AzTEC3 is located at a redshift of $z_\mathrm{spec}=5.298$, and that it is associated with a proto-cluster at the same redshift. Follow-up observations with the VLA and PdBI (targeting CO 2-1, 5-4, and 6-5; \cite{riechers2010}; \cite{riechers2014}) revealed a large molecular gas reservoir with a mass of $5.3\times10^{10}$ ($\alpha_\mathrm{CO}/0.8$)~\msol \, (where $\alpha_{\rm CO}$ is the CO luminosity-to-molecular gas mass conversion factor). Riechers et al.\ (2014) find a deconvolved [\ion{C}{II}] size (FWHM) of $3.9\times2.1$~kpc$^2$, 2.5~kpc  in dust continuum, and a dynamical mass of $9.7\times10^{10}$~\msol. Detailed modeling of the SEDs of the source (\cite{dwek2011}) yielded that the dust contained in the source likely formed over a period of $\sim200$ Myr with a SFR of $\sim500$~\msolyr \ and a top-heavy IMF (the SFR can rise up to 1800~\msolyr \ if other scenarios are assumed; see \cite{dwek2011} for details).
\item[$\bullet$] {\em J1000+0234} is a $\sim3\sigma$ detection in the JCMT/AzTEC-COSMOS map identified to lie at $z_\mathrm{spec}=4.542$ based on its Keck~II/DEIMOS spectrum (\cite{capak2008}), and further CO follow-up with the VLA and PdBI (\cite{schinnerer2008}). Capak et al.\ (2008) have shown that the Ly$\alpha$ line breaks into multiple, spatially separated components with an observed velocity difference of up to 380~km~s$^{-1}$ and that the  morphology of the source indicates a merger. Schinnerer et al.\ (2008) report a molecular gas mass of $2.6\times10^{10}$ ~\msol\ assuming a ULIRG-like conversion factor and a dynamical mass of $1.1\times10^{11}$~\msol\ assuming a merger scenario. 
\item[$\bullet$] {\em AzTEC/C159} is a $3.7\sigma$ detection in the ASTE/AzTEC-COSMOS 1.1~mm survey of the inner COSMOS 1~deg$^2$ (\cite{aretxaga2011})\ located at $\alpha_{2000.0}=09^{\rm h} 59^{\rm m}  29\fs75$, $\delta_{2000.0}=+01\degr55\arcmin35\arcsec$. Its deboosted 1.1~mm flux density is $3.3\pm1.3$~mJy (deboosted by a factor of $\sim1.5$). Here we associate it with a radio source within the AzTEC beam, $12\arcsec$ away from the centroid of the AzTEC detection at  $34\arcsec$ resolution. A DEIMOS spectrum for this source yields a spectroscopic redshift of $z_{\rm spec}=4.569$ (see \s{sec:specdata} ).  

\item[$\bullet$] {\em Vd-17871} has initially been identified as an Lyman-break galaxy (LBG) exhibiting faint radio emission (consistent with the properties of e.g.\ AzTEC1 and J1000+0234). Subsequently it was followed up in the (sub-)millimeter wavelength regime. It has been detected at 1.2~mm with MAMBO on the IRAM 30 m telescope (A.~Karim et al., in prep.). Keck~II/DEIMOS, VLT/VIMOS, and PdBI spectroscopy (A.~Karim et al., in prep.) have shown that Vd-17871 is located at a redshift of $z_\mathrm{spec}=4.622$, and that it breaks-up into two components, separated by 1\farcs5. Karim et al.\ show that both components are at the same (spectroscopic) redshift, suggesting a galaxy merger. 

\item[$\bullet$] {\em AK03} has initially been identified as an LBG exhibiting faint radio emission (consistent with the properties of, e.g., AzTEC1 and J1000+0234), and subsequently followed up in the (sub-)millimeter wavelength regime. It is a 3.7$\sigma$ detection in the SCUBA-2 450~$\mu$m map (\cite{casey2013}, their source 450.61), and a tentative, 3$\sigma$, detection in the JCMT/AzTEC map (similar to J1000+0234; \cite{scott2008}; M.~Yun, priv. comm.). The system, showing two components with a projected separation of $\sim0\farcs9$ in the optical, lies $1\farcs3$ away from the SCUBA-2 source, and $7\arcsec$ away from the position of the AzTEC source. Given the $\sim 18\arcsec$ (FWHM) JCMT/AzTEC beam, the positional accuracy of the tentative mm-source is $\sim 6\arcsec$, hence not ruling out an association with the radio source (associated with the southern component, for which a spectroscopic redshift is not available). A Keck~II/DEIMOS spectrum of the Northern component of AK03 (AK03-N hereafter) puts it at a redshift of $z_\mathrm{spec}=4.747$, while the Southern component (AK03-S hereafter), detected also in the UltraVISTA near-IR and radio maps, has a comparable photometric redshift. Following the photometric-redshift computation described in detail in Ilbert et al.\ (2009) for AK03-S (coincident with radio emission; see Fig.~\ref{fig:stamps}) we find a photometric redshift of $z=4.40\pm 0.10$ if galaxy templates (\cite{ilbert2009}) are used or $z=4.65\pm 0.10$ if quasar templates (\cite{salvato2009}) are employed. The $\chi^2$ statistics suggest a roughly equally good fit for the galaxy and quasar libraries, and the errors on the photometric redshift reported reflect the uncertainties obtained via a comparison with spectroscopic redshifts. Thus, the two (North and South) components of AK03 may be associated, and perhaps merging.

\end{itemize}

\section{Multi-wavelength properties and data}
\label{sec:data}

COSMOS is an imaging and spectroscopic survey of an equatorial 2 deg$^2$\ field (\cite{scoville2007}). It provides all ancillary data needed for a detailed interpretation of our $z>4$ SMG sample given the deep ground- and space-based photometry in more than 30 bands, from the X-rays to the radio (\cite{scoville2007}). The IR to radio photometric measurements for our SMGs are reported in Tables~\ref{tab:photherschel},~\ref{tab:photmm}, and~\ref{tab:radio}, and the multi-wavelength data are described in more detail below. 

\subsection{X-ray properties}

The SMGs in our sample are located in an area of the COSMOS field covered by deep ($\sim160-190$ ks) {\it Chandra} images, which 
are part of the C-COSMOS (\cite{elvis2009}) and {\it Chandra} COSMOS Legacy surveys (XVP; F.~M.~Civano et al., in prep.). 
None of our SMGs is detected in X-rays. On the basis of the sensitivity of the above surveys (see \cite{puccetti2009}), 
the detection limit at the position of the sources is $\sim3-5\times 10^{-16}$~${\rm~erg~cm}^{-2}~{\rm s}^{-1}$ in the $0.5-2$ keV band, 
which corresponds to an upper limit on the rest-frame luminosity of $10^{43.5}$~erg~s$^{-1}$ in the same band, computed assuming $z=4.5$, a 
power law model with spectral index $\Gamma=1.4$ and Galactic hydrogen column density $N_{\rm H,\,Gal}=2.6 \times 10^{20}$ cm$^{-2}$ and no intrinsic absorption. 
We also performed an X-ray stacking analysis, which does not yield a detection even for the stacked signal. This sets an upper limit on the X-ray luminosity to 10$^{43.1}$ erg~s$^{-1}$. We note that this limit is significantly higher than the expected X-ray emission from SMGs (\cite{laird2010}) if their X-ray emission were due only to star formation.

\subsection{UV-FIR data}

The UV-mid-IR photometry is taken from the most recent COSMOS photometric catalog, with UltraVISTA ($Y,\,J,\,H,\,K_{\rm s}$) DR1 photometry added (see \cite{smolcic2012}; \cite{ilbert2013}; \cite{mccracken2012}; \cite{capak2007}). 

To assure the most accurate FIR measurements for our sources, we use deep Photodetector Array Camera and Spectrometer (PACS) 100 and 160$\,\mu$m and Spectral and Photometric Imaging Receiver (SPIRE) 250, 350 and 500$\,\mu$m observations of the COSMOS field provided by the \textit{Herschel} Space Observatory (\cite{pilbratt2010}). The PACS observations were taken as part of the PACS Evolutionary Probe (PEP\footnote{http://www.mpe.mpg.de/ir/Research/PEP}; \cite{lutz2011})
 guaranteed time key program, while the SPIRE observations were taken as part of the \textit{Herschel} Multi-tiered Extragalactic Survey 
 (HerMES\footnote{http://hermes.sussex.ac.uk}; \cite{oliver2012}).
We extract the \textit{Herschel} PACS and SPIRE photometry directly from the maps via a point source function (PSF)-fitting analysis (\cite{magnelli2013})
 using as prior positions the interferometric position of our SMGs and, where required, the position of MIPS 24$\,\mu$m sources blended with our SMGs.
The \textit{Herschel} photometry is presented in Table~2 where we also report the reliability of the \textit{Herschel} fluxes. 
We note that we cannot decompose the FIR emission for SMG components separated by $\lesssim 1\arcsec$, and thus for those systems we report the integrated photometry. 

\subsection{Radio data}

The 1.4~GHz (20~cm) radio photometry for our SMGs is taken from  the VLA-COSMOS 1.4~GHz Large and Deep Projects (\cite{schinnerer2007}, 2010). These reach an rms of $\sim7-12~\mu$Jy~beam$^{-1}$ over the 2~deg$^2$\ COSMOS field. The 20~cm fluxes for our sources are listed in Table~3. 
Our sources are not detected in the 324~MHz (90 cm) VLA-COSMOS map at  $8\farcs0 \times 6\farcs0$ angular resolution (\cite{smolcic2014}), as expected given an rms of 0.5 mJy~beam$^{-1}$ in the 324~MHz map. However, all sources but J1000+0234 are detected in the deep GMRT-COSMOS 325~MHz map (at a resolution of $10\farcs8 \times9\farcs5$ and reaching an rms of $70-80$ $\mu$Jy~beam$^{-1}$ around our SMGs; A.~Karim et al., in prep.), and their flux densities (based on the peak flux values) are reported in the last column of \t{tab:aztecphot} . The reduction and imaging of the GMRT-COSMOS 325 MHz map is described in detail in Karim et al.\ (in prep.).

We here also use 3~GHz data recently taken with the upgraded,  Karl G. Jansky Very Large Array (VLA; VLA-COSMOS 3~GHz Large Project, Smol{\v c}i{\'c} et al., in\ prep.). 
The data reduction and imaging of the sub-mosaic of the full COSMOS field is described in detail in Novak et al.\ (2014). The sub-mosaic is 
based on 130 hours of the observations taken in A-  and C-configurations (110 and 20 hours, respectively) and contains the pointings (P19, P22, P30, P31, P39, and P44) targeting our six SMGs closest to the pointing phase center. 
The imaging of the pointings was performed using the CASA task CLEAN 
using the MFMS option 
that images the full bandwidth simultaneously (\cite{rau2011}) and simultaneously preserves the best possible resolution ($0\farcs6\times0\farcs7$) and the best possible rms ($\sim4.5$ $\mu$Jy~beam$^{-1}$) in the map (see \cite{rau2011} for details). The 3 GHz radio stamps of our SMGs at $0\farcs6\times0\farcs7$ resolution are shown in Fig.~2. All sources are detected in our 3 GHz radio mosaics. 
We derive their integrated fluxes using a 2D Gaussian fit to the sources (via the AIPS task JMFIT) and peak fluxes using the AIPS task MAXFIT. The ratio of the integrated and peak fluxes for a given radio source is a direct measure of the extent of the source (e.g. \cite{bondi2008}; \cite{kimball2008}). Based on this, we find that 
at $0\farcs6\times0\farcs7$ resolution AzTEC/C159 is clearly resolved, J1000+0234 and AK03 may be marginally resolved, while AzTEC1, AzTEC3, and Vd-17871 remain unresolved. We note that although our analysis suggests extended or multiple-component emission for J1000+0234 and AK03, their integrated fluxes obtained via the AIPS task TVSTAT down to the $2\sigma$ level are consistent with their peak fluxes, such that we cannot unambiguously determine whether the sources are really resolved at this resolution. Higher resolution radio imaging is required to unambiguously determine the radio-emitting sizes for these sources. In Table~\ref{tab:radio} we report the deconvolved 3~GHz sizes of our sources. For all the unresolved sources including the two marginally resolved ones, the total flux is then set equal to the peak brightness, while for AzTEC/C159 we use the AIPS task TVSTAT to infer the total flux within the region outlined by the $2\sigma$ contour. The fluxes are reported in Table~\ref{tab:aztecphot}.

\begin{figure}[!h]
\centering
\resizebox{\hsize}{!}{
\includegraphics{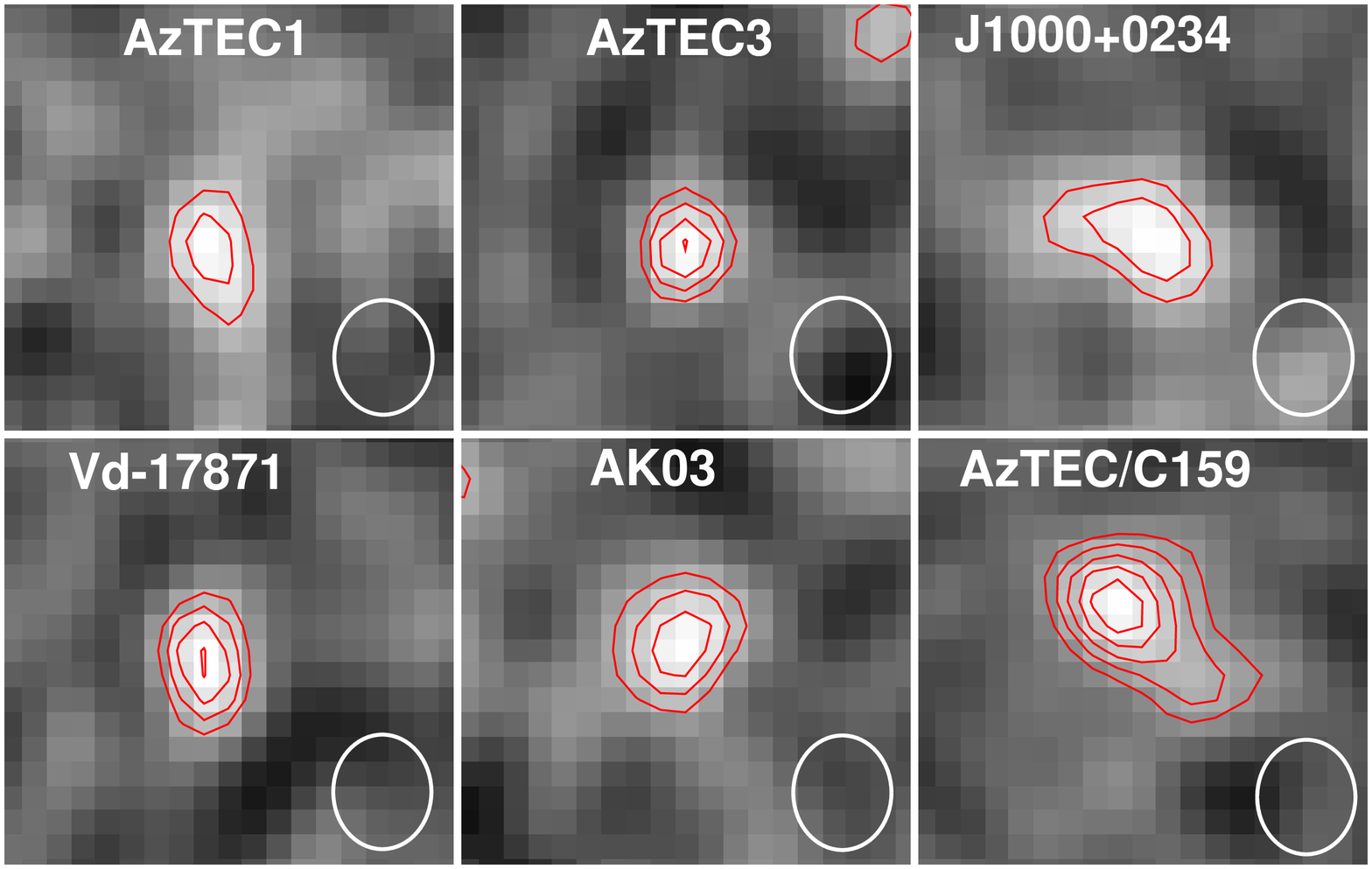}
}
\caption{The VLA 3 GHz (10 cm) images ($\sim 3''\times 3''$ in 
size) of the studied SMGs. The overlaid red contours show the emission levels 
starting from $2\sigma$ and increasing in steps of 
$1\sigma=4.5$ $\mu$Jy~beam$^{-1}$  (the corresponding negative contours levels are shown by dashed contours). The synthesized beam size, 
$0\farcs6 \times 0\farcs7$, is indicated in the bottom right corner of each 
panel.}
\label{figure:3GHz}
\end{figure}

\subsection{Spectroscopic data}
\label{sec:specdata}

All sources in the sample were targeted with DEIMOS on Keck~II. In \f{fig:spec} \ we show the 1D spectra for 
AzTEC/C159 and AK03 (the spectra of the remaining sources are given in \cite{capak2008}; \cite{smolcic2011}; \cite{capak2011}, and A.~Karim 
et al., in prep. for J1000+0234, AzTEC1, AzTEC3, and Vd-17871, respectively). 
The spectra were taken early 2009 under clear weather conditions and
$\sim1\arcsec$ seeing and a 4 hr integration time split into 30 minute 
exposures using the 830 lines mm$^{-1}$ grating tilted to $7\,900 \,\AA$ and the 
OG550 blocker. The objects were dithered $\pm3 \arcsec$ along the slit to 
remove ghosting. The data reduction was performed via the modified DEEP2 DEIMOS
pipeline (see \cite{capak2008}). In both spectra Ly$\alpha$ is clearly detected in emission setting the redshifts to $z=4.747$ and $z=4.569$ for AK03 and AzTEC/C159, respectively. In \f{fig:spec} \ we also show QSO (\cite{croom2002}) and LBG composite spectra constructed by Shapley et al. (2003) from almost 1000 LBGs. The continuum is barely detected for AK03 and AzTEC/C159, and the width of the Ly$\alpha$ line rules out powerful AGN contribution to the SMGs, consistent with their X-ray properties, yet the contribution of a less powerful or obscured AGN cannot be ruled out.

\begin{figure}[ht!]\vspace{-1cm}
\includegraphics[bb= -50 10 654 754, width=\columnwidth, angle=90]{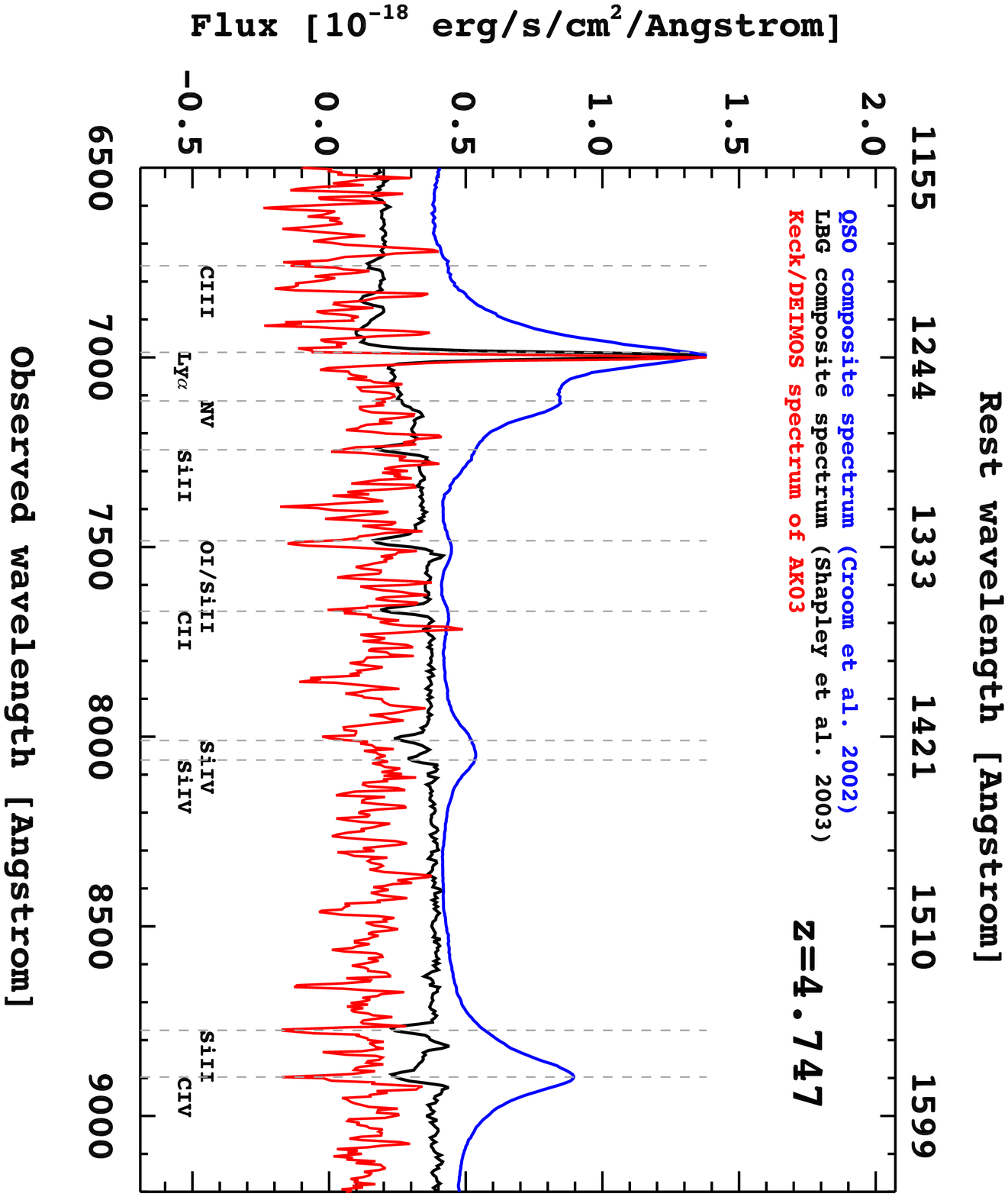}\\
\includegraphics[bb= -50 10 654 754, width=\columnwidth, angle=90]{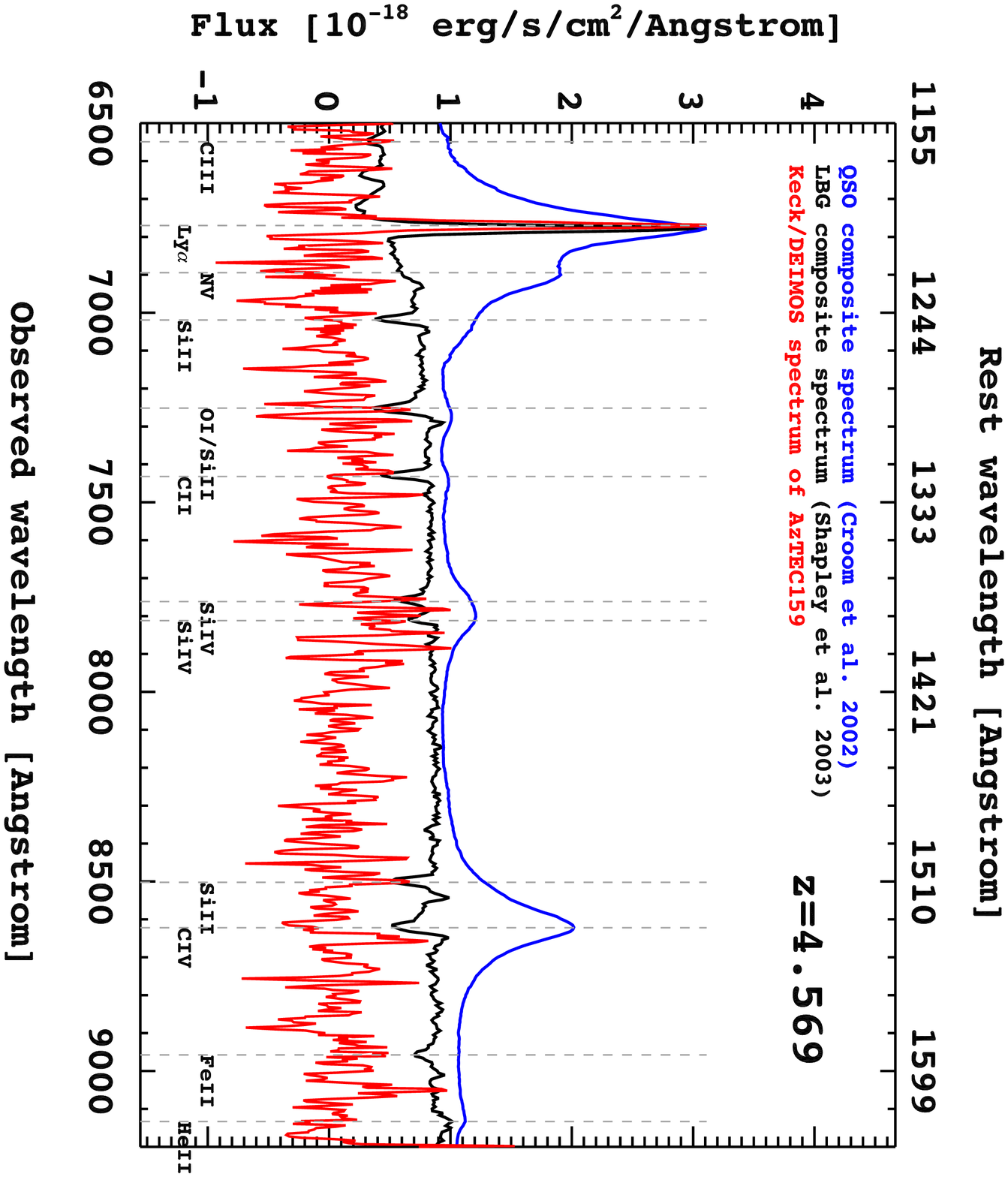}
\caption{1D Keck~II/DEIMOS spectra (red) for AK03 (top panel), and AzTEC/C159 (bottom panel). Also shown are QSO (blue) and LBG (black) composite spectra at the given redshift (labels in the panels). The vertical dashed lines indicate 
positions of various spectral lines, indicated in the panel. The spectroscopic redshift of the SMG is indicated in the top right corner of each panel. 
\label{fig:spec}}
\end{figure}

\section{Spectral energy distributions and derived properties}
\label{sec:results}

In this section we analyze the NUV to radio SED properties of our sample of $z>4$ SMGs. We self-consistently fit the NUV-mm SEDs of our SMGs using the MAGPHYS model package (\s{sec:sedmagphys} ), we then provide an independent fit of the dust SEDs using FIR data only (\s{sec:sedfir} ), and finally analyze the radio SEDs of our sources (\s{sec:sedradio} ).

\subsection{NUV-mm SED fitting}
\label{sec:sedmagphys}

Here we analyze the NUV-mm SEDs of our SMGs via the MAGPHYS (Multi-wavelength Analysis of Galaxy Physical Properties) model package (\cite{dacunha2008}), which self-consistenly balances dust attenuation in the UV-NIR and dust heating in the IR. The UV-NIR SED is modeled based on the Bruzual \& Charlot (2003) and Charlot \& Fall (2000) stellar population synthesis models. The code then further calculates the absorption of starlight by dust in stellar birth clouds and the  diffuse interstellar medium, and computes the SED of the power reradiated by this dust as a sum of three components: i) polycyclic aromatic hydrocarbons (PAHs), ii) hot dust grains (130-250~K)  characterizing the MIR continuum, and iii) grains in thermal equilibrium with temperature in the range of 30-60~K. 

The MAGPHYS model package provides a library of model spectra that are fit (via a standard $\chi^2$ minimization procedure) to the observed SED. Here we use a library optimized for starburst (ULIRG) galaxies (\cite{dacunha2010}). We, however, also utilize a second library, optimized for normal galaxies (\cite{dacunha2008}) in the case of AK03-N and -S as the ULIRG library  cannot adequately fit the sources' SEDs (see below). 

\begin{figure*}[t!]
\includegraphics[bb =  74 560 558 720, scale=0.6]{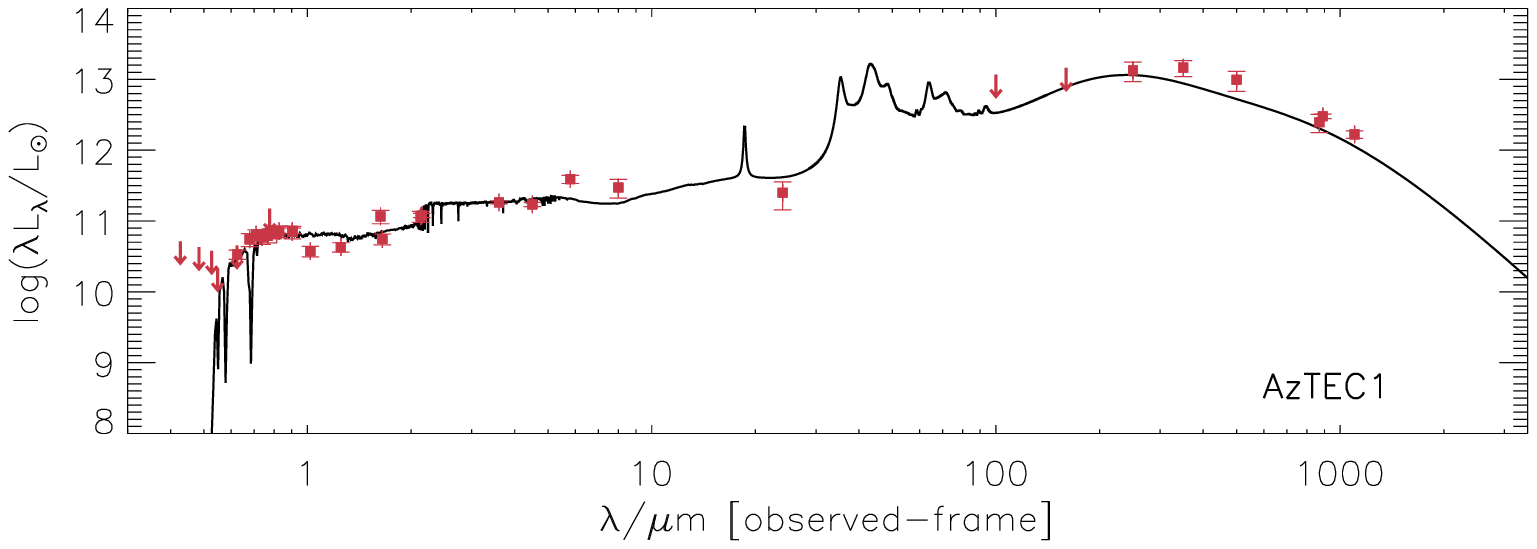}
\includegraphics[bb =  104 560 558 720, scale=0.6]{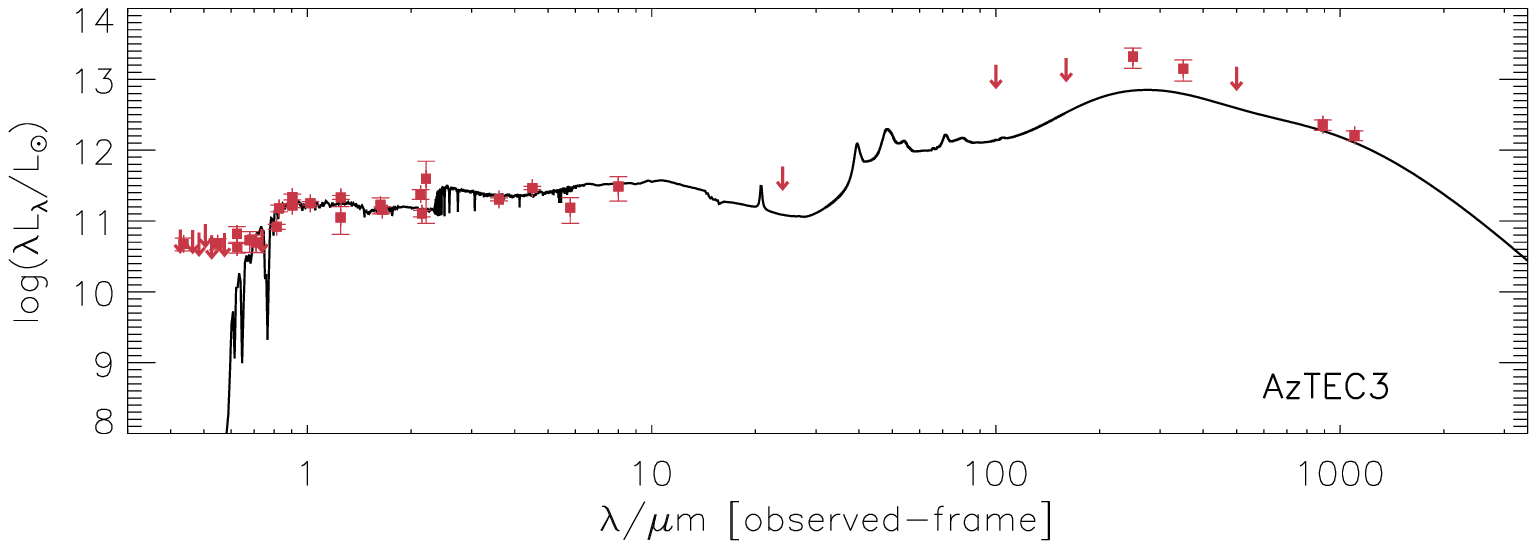}\\
\includegraphics[bb =  74 560 558 720, scale=0.6]{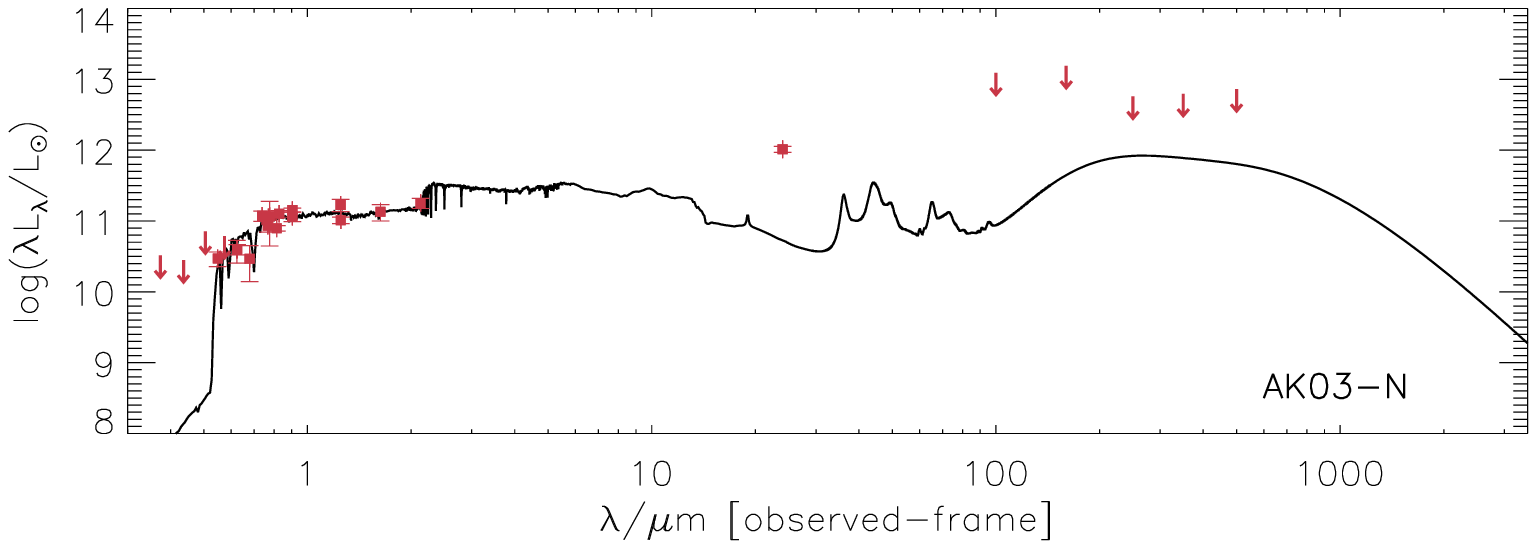}
\includegraphics[bb =  104 560 558 720, scale=0.6]{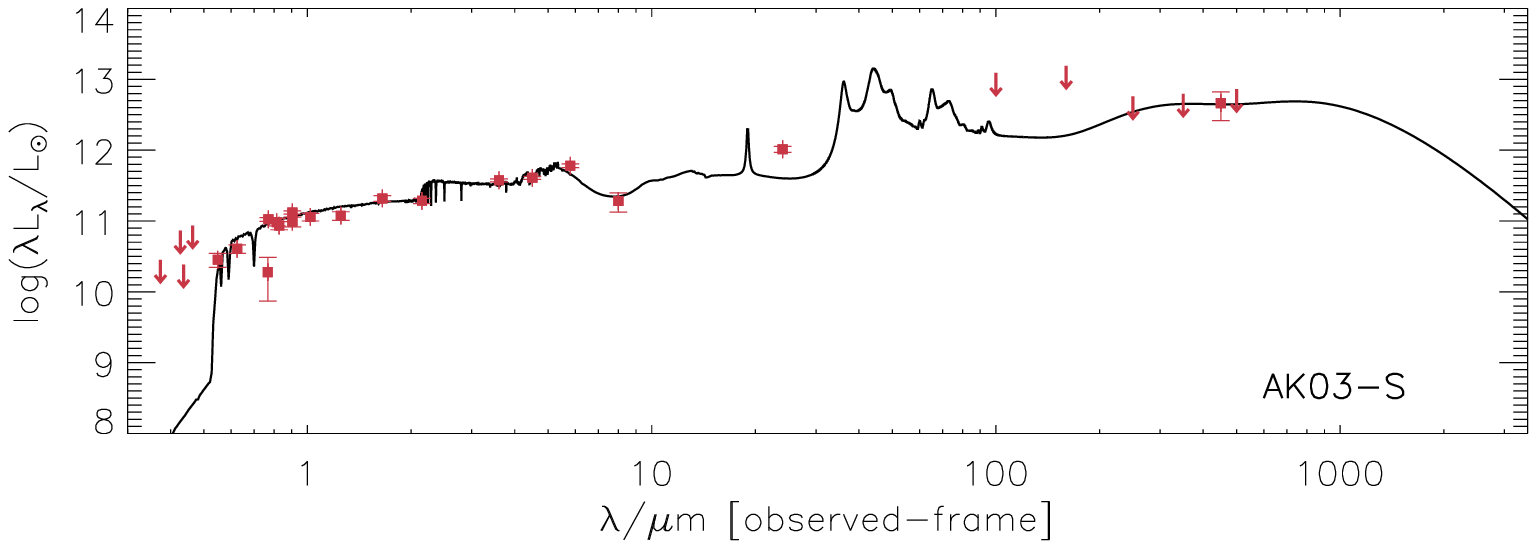}\\
\includegraphics[bb =  74 560 558 720, scale=0.6]{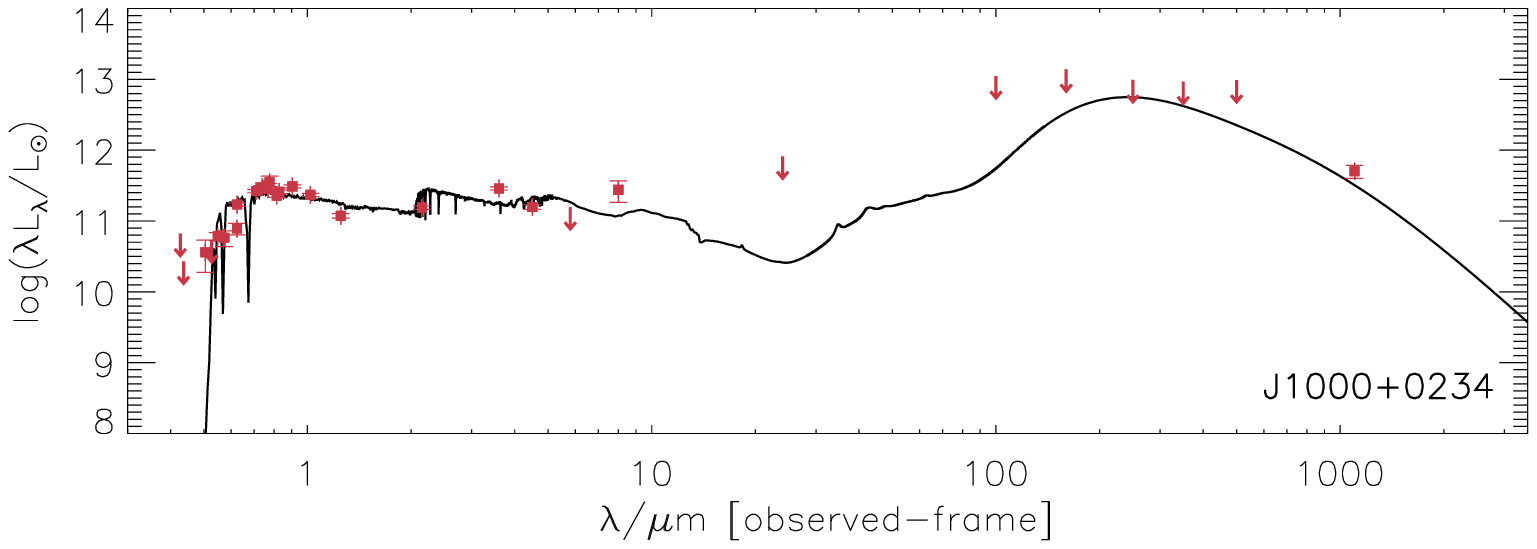}
\includegraphics[bb =  104 560 558 720, scale=0.6]{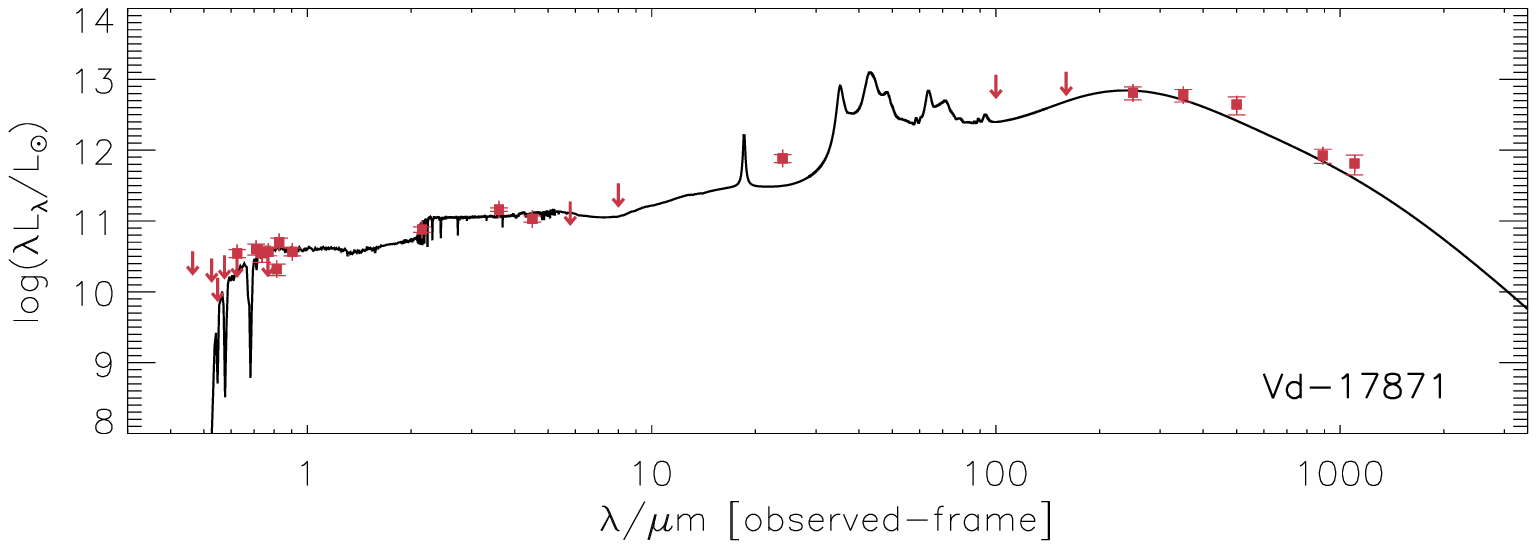}\\
\includegraphics[bb =  74 560 558 720, scale=0.6]{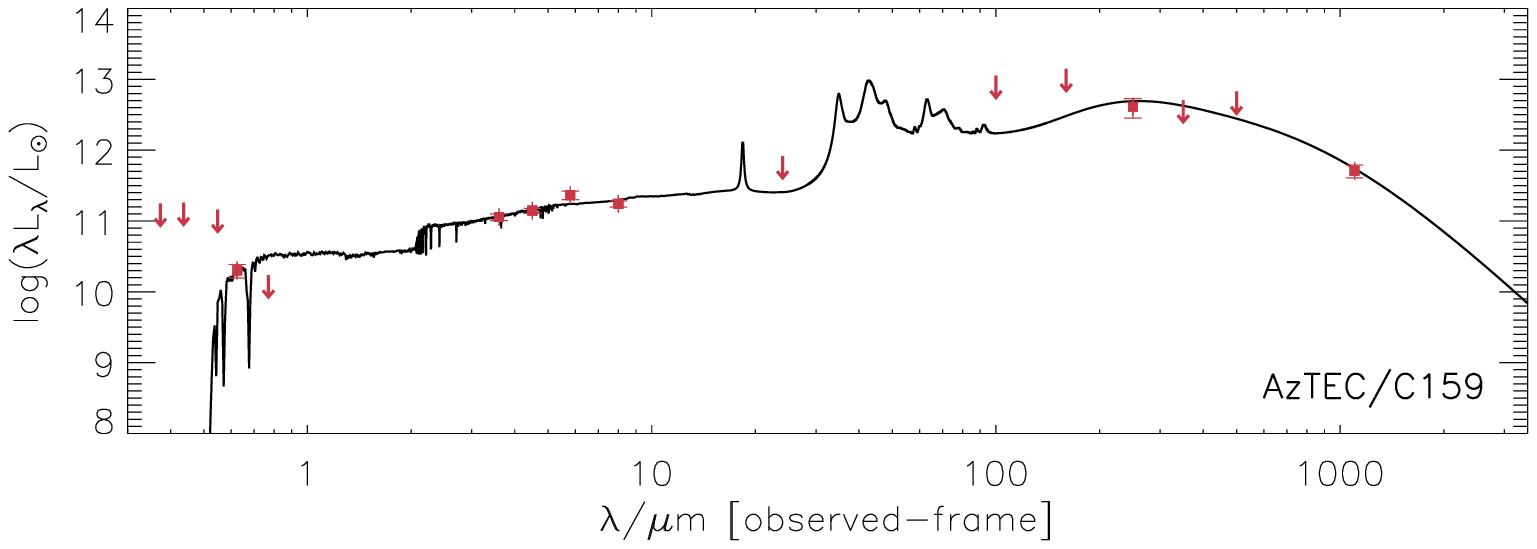}
\caption{Optical to IR SED fits to our $z>4$ SMGs (points for detections, arrows for $4\sigma$ upper limits) using the MAGPHYS 
routine (full black line). 
 \label{fig:Optsed}}
\end{figure*}

We perform the fitting for each of our SMGs using their full NUV-mm photometry (total fluxes de-reddened for foreground Galactic extinction excluding the intermediate band containing the Ly$\alpha$ emission line)  extracted at the position of the  optical/UltraVISTA counterpart closest to the spectroscopic/mm source (see Fig.~\ref{fig:stamps}). The SEDs and the best-fit models for all of our SMGs are shown in Fig.~\ref{fig:Optsed}, and 
in Table~\ref{tab:magphys} we list the following parameters from the best-fit models to the observed SEDs for our $z>4$ SMGs:  
the age of the oldest stars in the galaxy giving a formation timescale ($t_\mathrm{form}$), total $V$-band optical depth of the dust seen by young stars in their birth clouds ($\tau_\mathrm{V}$), and the stellar mass ($M_{\star}$). For completeness we also list the total dust luminosity ($L_\mathrm{dust}$) calculated over the wavelength range 3--1\,000 $\mu$m, and dust mass ($M_\mathrm{dust}$). However, in the following (Sect.~4.1.2) we independently determine these quantities based on modified black-body and Draine \& Li (2007, hereafter DL07) models, and we adopt the last as the reference value for further analysis. In particular, the IR luminosities derived from the DL07 model refer to the commonly adopted rest-frame wavelength range $8-1\,000~\mu$m, allowing us to make a direct comparison with SMGs' IR luminosities reported in the literature (Sect.~5).

\subsection{FIR properties}
\label{sec:sedfir}

To infer the IR luminosities (in the range $8-1\,000~\mu$m) and dust masses of our $z>4$ SMGs we fit their MIR-mm SEDs  using an optically thin ($\tau \ll 1$) modified 
 black-body, and the DL07 dust model.
The latter model describes the interstellar dust as a mixture of carbonaceous and amorphous silicate grains and has four free parameters:
(i) $q_{{\rm PAH}}$, which controls the fraction of dust mass in the form of PAHs;
(ii) $\gamma$, which controls the fraction of dust mass exposed to a power-law ($\alpha=2$) radiation field (relative to the local interstellar value) ranging from $U_{{\rm min}}$ to $U_{{\rm max}}$; the rest of the dust mass (i.e., $1-\gamma$) being exposed to a radiation field with a constant intensity $U_{{\rm min}}$;
(iii) $U_{{\rm min}}$, which controls the minimum radiation field seen by the dust ($U_{{\rm max}}$ is fixed to a value of $10^6$); and
(iv) $M_{\rm dust}$, which controls the normalization of the SED.
Following the prescriptions of DL07, we built a grid of models with different PAH abundances ($0.47\%<q_{{\rm PAH}}<4.6\%$), values of $U_{{\rm min}}$ ($0.7-25$) and values of $\gamma$ ($0.0-0.3$).
The best-fit model to each SED, and the corresponding $1\sigma$ error  is then found via  standard $\chi^2$ minimization.
The fits are shown in \f{fig:IRsed} \ and the results are summarized in 
Table~\ref{tab:ir}. From the inferred IR luminosities we then further estimate the IR-based SFRs for 
our $z>4$ SMGs, also reported in Table~\ref{tab:ir}, using the standard 
$L_{\rm IR}$-to-SFR conversion of Kennicutt (1998), assuming a Chabrier IMF, 
${\rm SFR\,[M_{\odot}\,yr^{-1}]\,=\,10^{-10}\,}L_{{\rm IR}}\,{\rm [L_{\odot}]}$.

As expected, the $M_{\rm dust}$ values derived through modified blackbody fits are systematically lower by a factor of $\sim3-4$ than those from 
the DL07 dust model, because single temperature models do not take into account warmer dust emitting at shorter wavelengths (\cite{magnelli2012}). 
We also find that in general the dust luminosities and masses derived via MAGPHYS utilizing the NUV-mm SED and those based only on the FIR SED are mostly 
in reasonable agreement. We note that the results for J1000+0234 are to be taken with caution as the \textit{Herschel} photometry is affected by confusion, 
which is reflected in the large error bars. 
Furthermore, the large difference in dust mass for AK03-S derived via MAGPHYS and IR SED fitting might possibly arise due to the 
high optical thickness assumed in the MAGPHYS model, while the IR modified black-body fitting was performed assuming an optically-thin regime.

\begin{figure*}
\begin{center}
\includegraphics[bb =  60 25 504 320, scale=0.5]{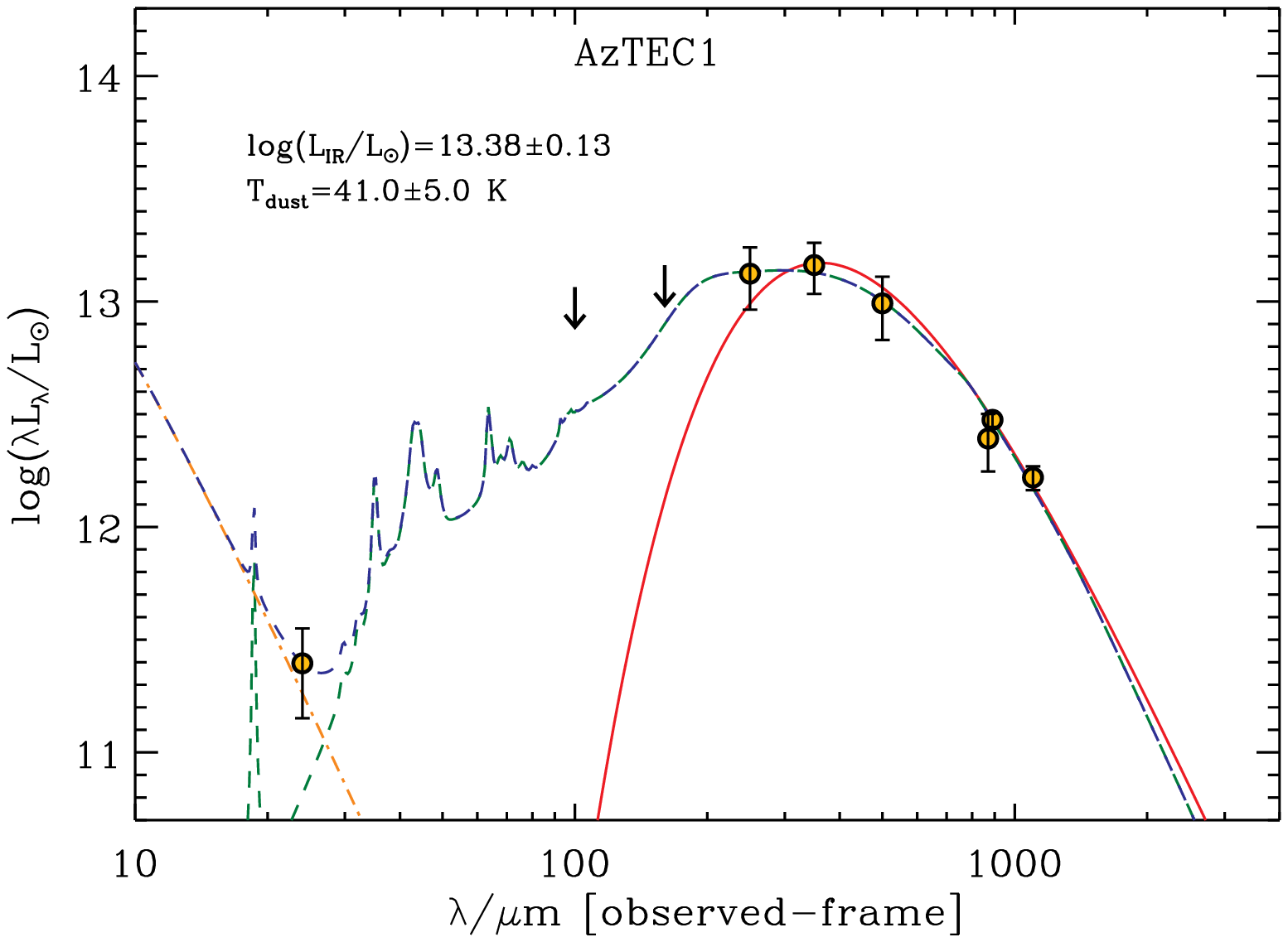}
\includegraphics[bb =  40 25 504 320, scale=0.5]{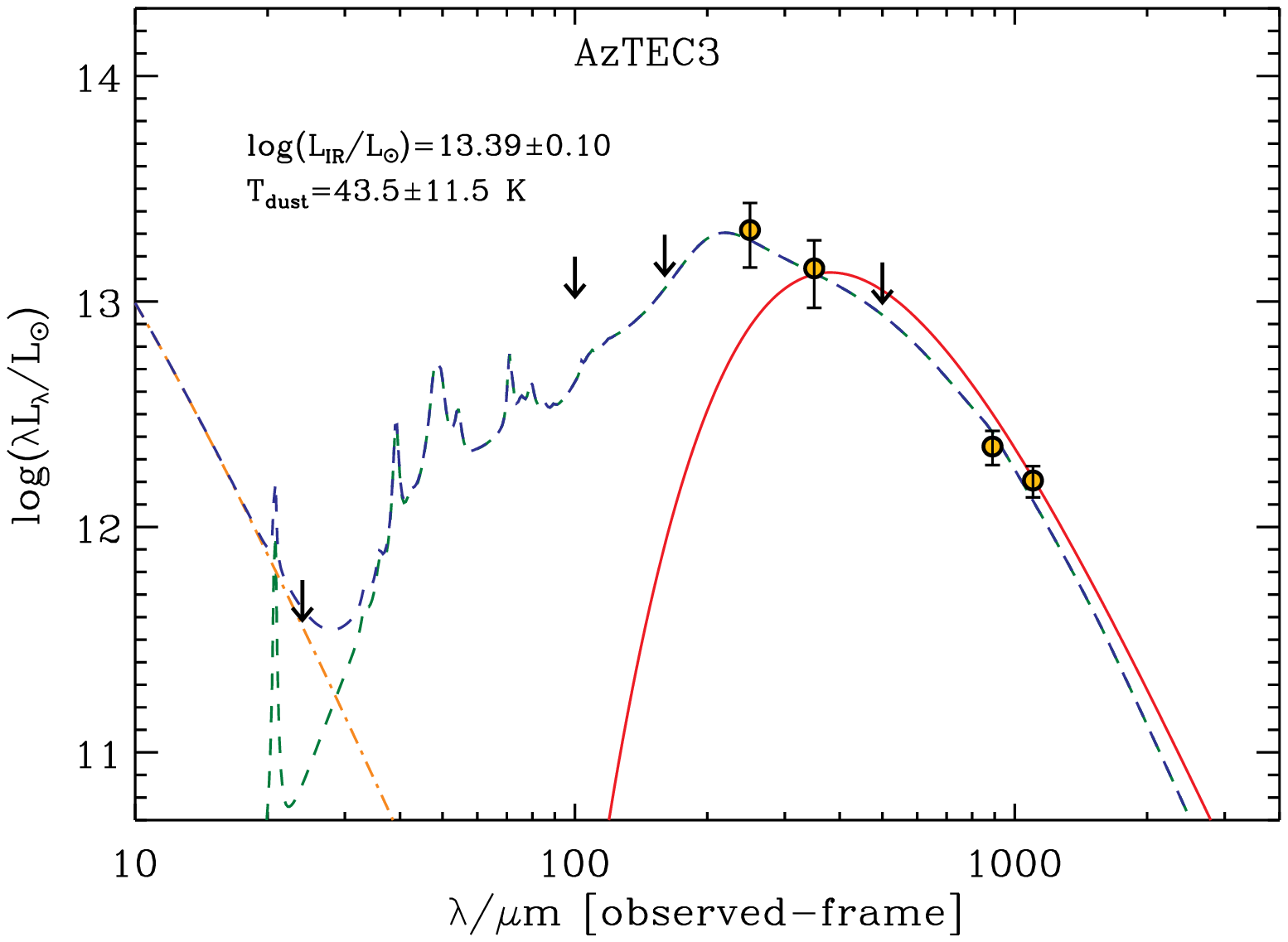}\\
\includegraphics[bb =  60 25 504 360, scale=0.5]{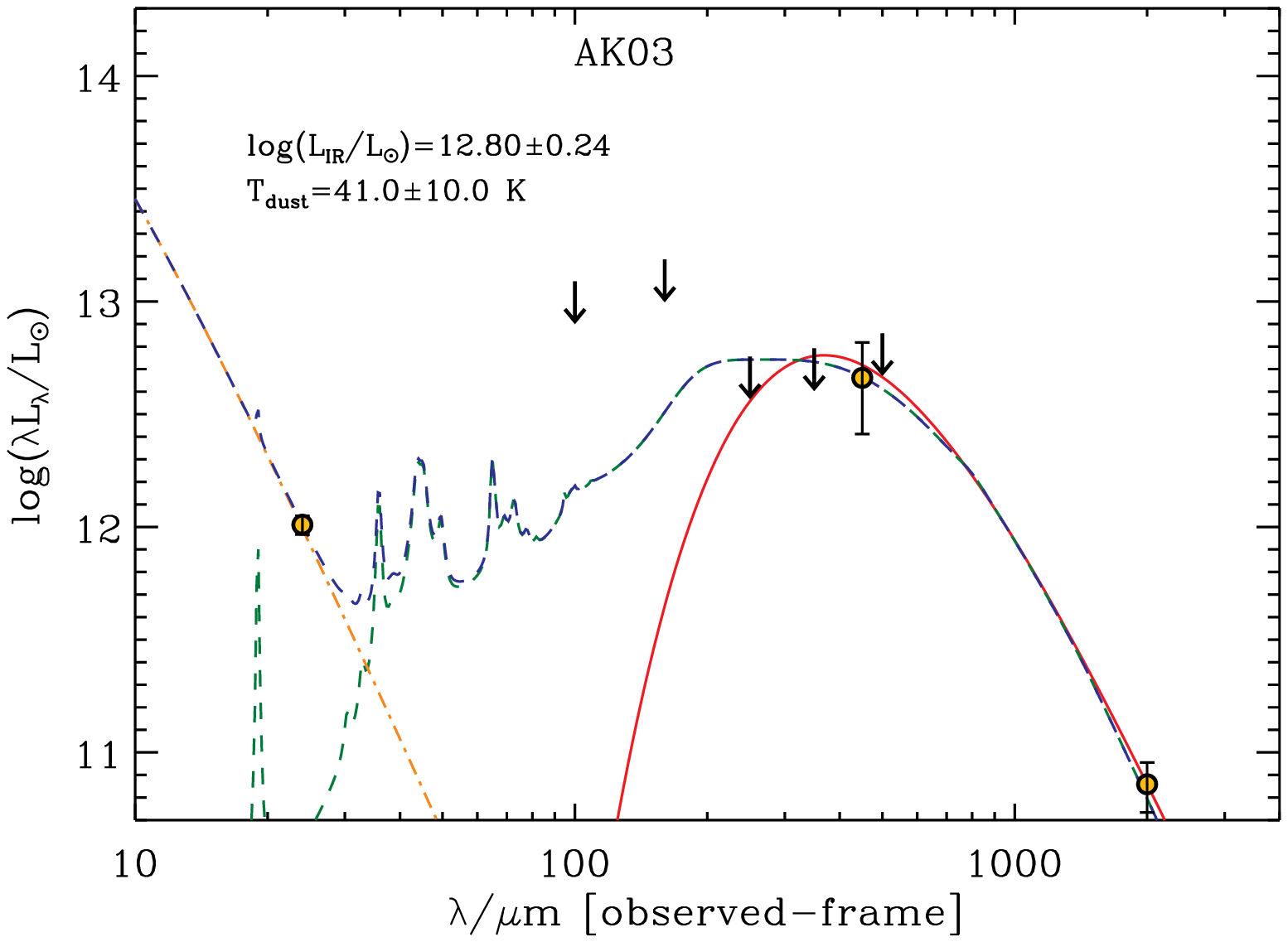}
\includegraphics[bb =  40 25 504 360, scale=0.5]{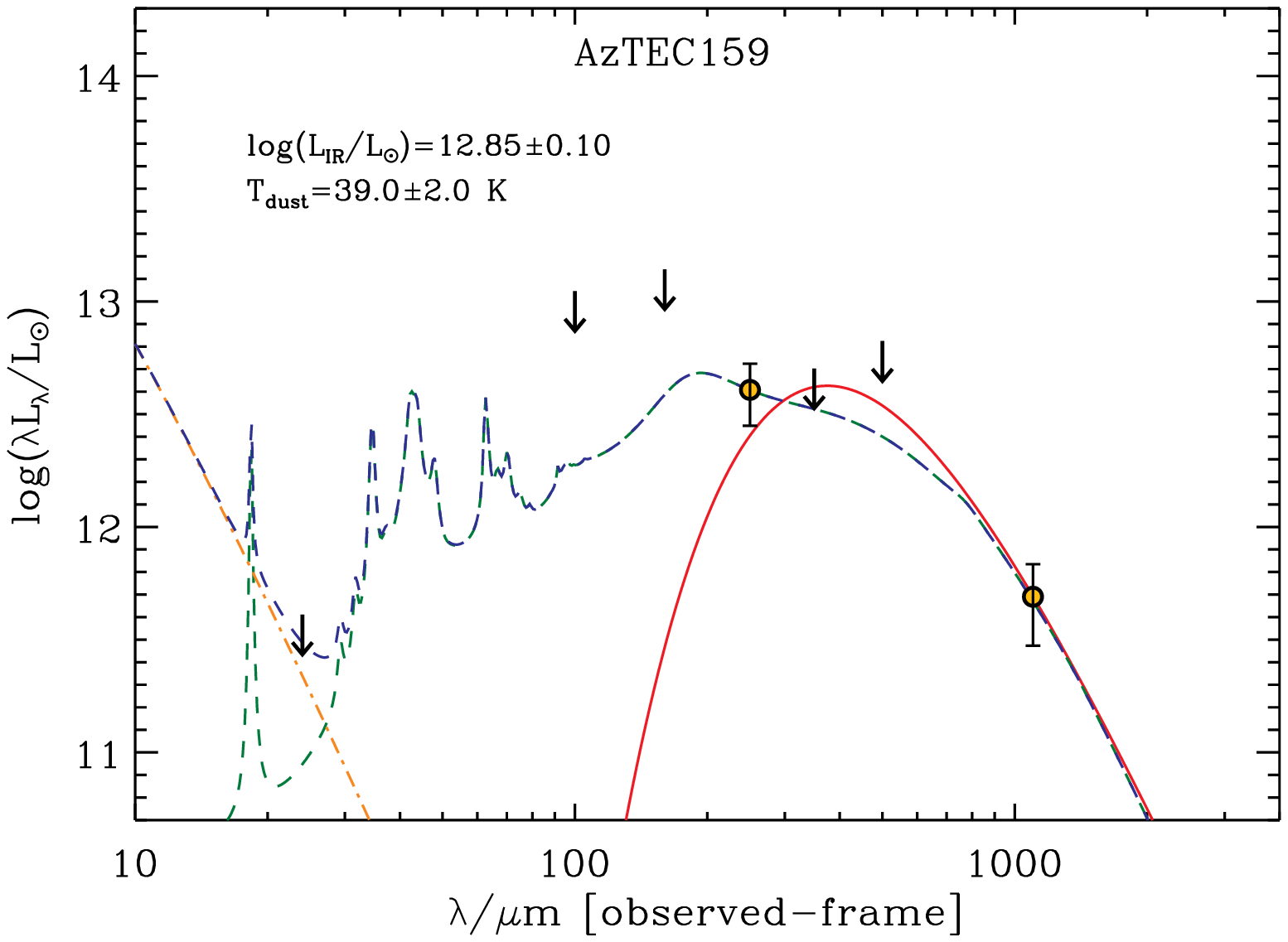}\\
\includegraphics[bb =  60 00 504 360, scale=0.5]{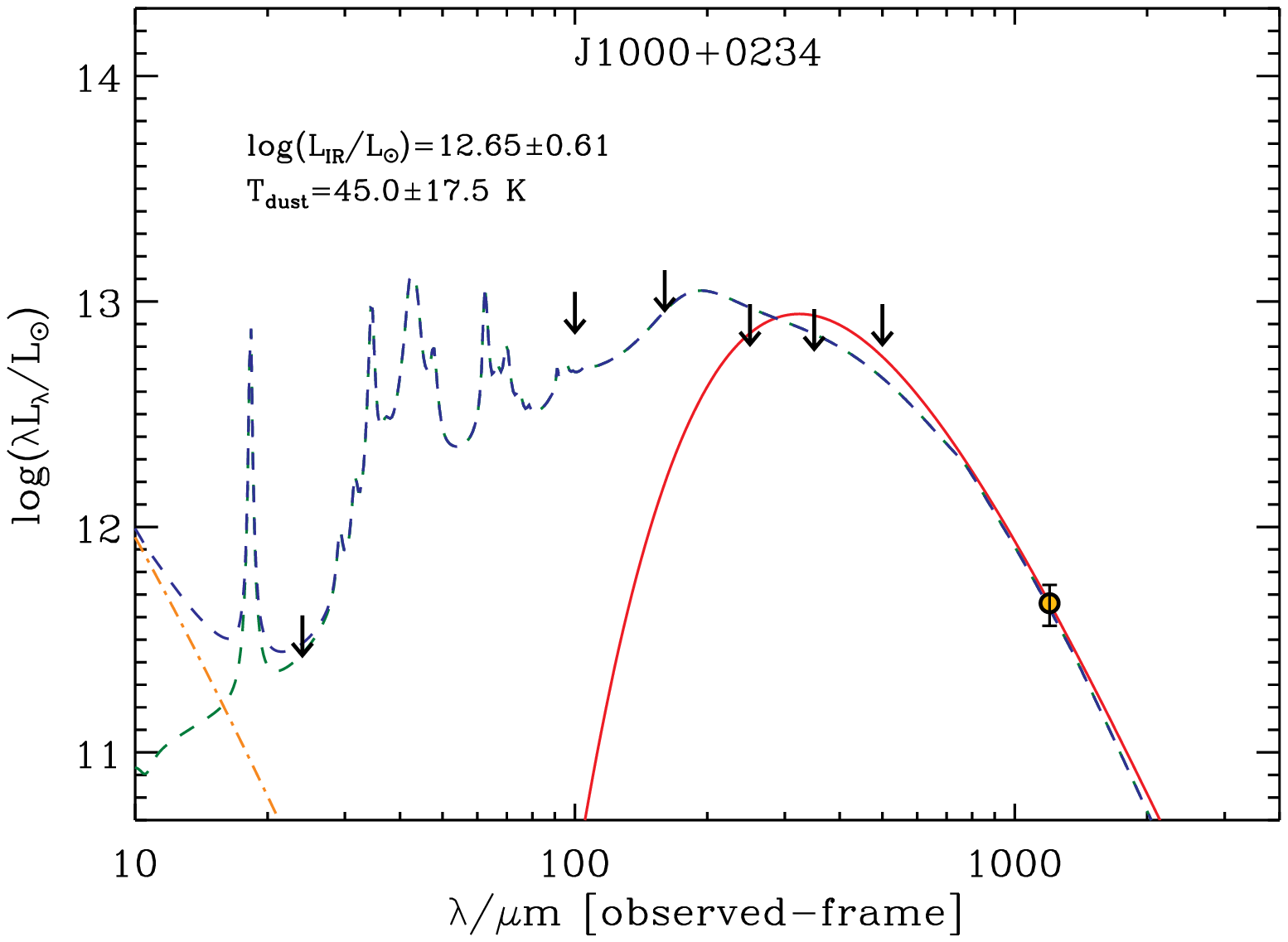}
\includegraphics[bb =  40 00 504 360, scale=0.5]{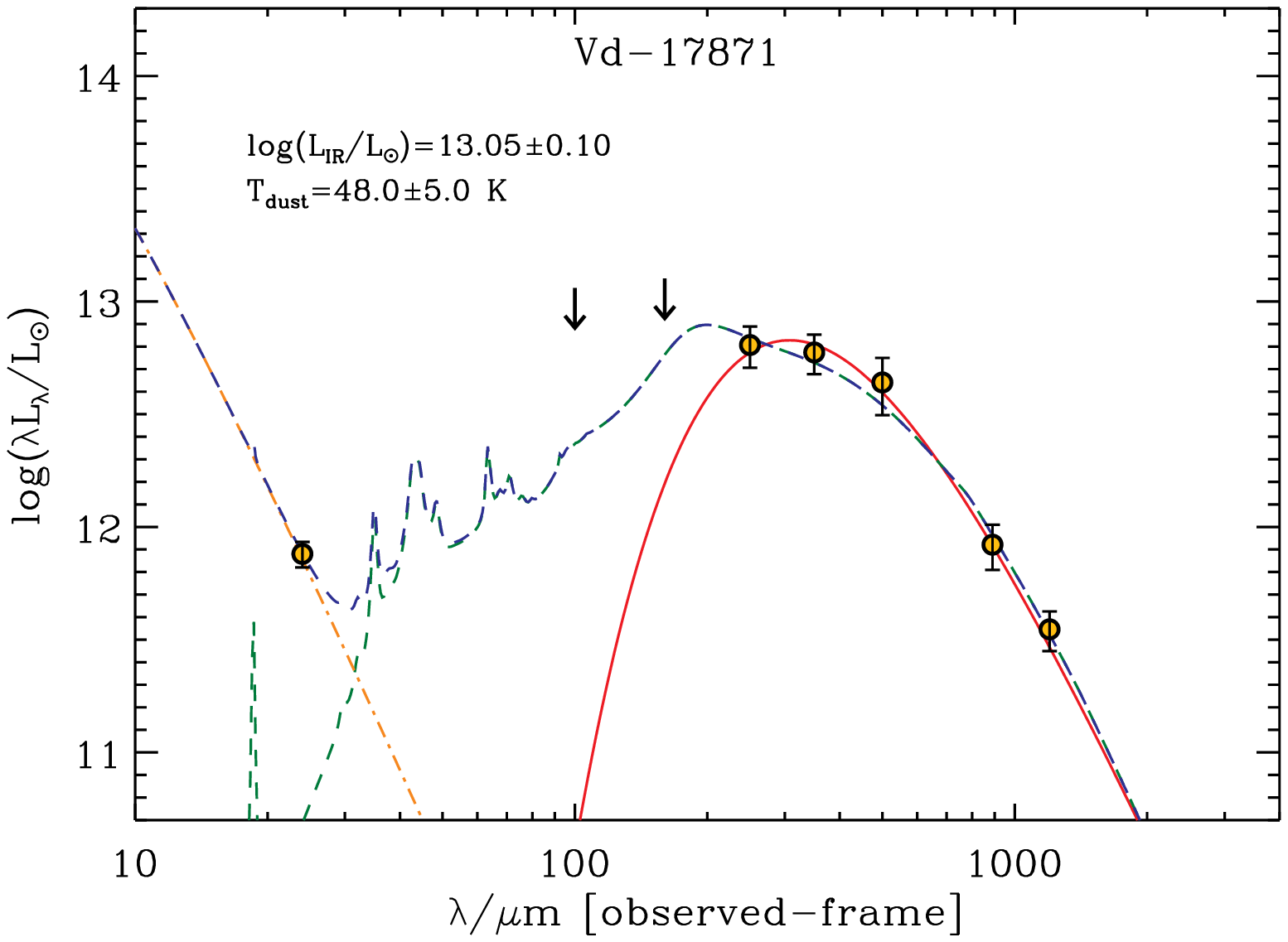}
\caption{The broadband SEDs of the studied SMGs. The downward pointing arrows 
denote upper limits to the corresponding flux densities. The red line 
represents a modified blackbody fit to the FIR-to-mm data points. The blue dashed line is the best fit model, obtained from the superposition of the best fit Draine \& Li (2007) model (long-dashed green line) and the stellar black body function (orange dash-dotted line) assuming a temperature of 5,000~K, scaled to the residual 24~$\mu$m ``stellar'' flux (i.e.\ observed 24~$\mu$m flux minus that obtained from the DL07 best-fit model). 
 \label{fig:IRsed}}
\end{center}
\end{figure*}

\subsection{Radio properties}
\label{sec:sedradio}

For our $z>4$ SMGs at the observed frequencies of 325~MHz, 1.4~GHz, and 3~GHz we are probing rest-frame frequencies of approximately 1.8~GHz, 7.7~GHz, and  16.5~GHz (assuming $z=4.5$). While the first frequency is expected to be in the region of the spectrum still unaffected by synchrotron self-absorption and energy losses, the latter two frequencies may be affected by energy losses. To account for the possible change in the spectral slope we derive the spectral indices separately at the high and low frequency ends taking the 1.4~GHz data point as the middle, reference point. The results are listed in Table~\ref{tab:radio}. 

The synchrotron spectrum for all sources is shown in \f{fig:synch} . Except AzTEC3 and AK03, it is consistent with a single power-law. 
The steepening of the synchrotron spectrum of AK03 at the high-frequency end may indicate an older age of the electrons exhibiting synchrotron emission. This would be consistent with the formation timescale of $\sim700$ Myr obtained for AK03-S from the UV-mm fitting via MAGPHYS (see Table~\ref{tab:magphys}). The synchrotron spectrum of AzTEC~3  flattens at the high-frequency end. As AzTEC~3 is the highest-redshift SMG in our sample ($z=5.299$) it is possible that the observed radio emission at these high rest-frame frequencies arises from thermal free-free emission (see, e.g.,\ Fig.~1 in \cite{condon1992}). 

We further compute the rest-frame monochromatic radio luminosity at 1.4~GHz ($L_{\mathrm{1.4GHz}}$) using the observed 325~MHz flux ($S_\mathrm{325MHz}$) as it is the closest to 1.4~GHz in the rest-frame, and thus requires the smallest (and hence least uncertain) K-correction:

\begin{equation}
L_{\mathrm{1.4GHz}} = \frac{4\pi D_\mathrm{L}^2}   {(1+z)^{1-\alpha}} S_\mathrm{325MHz} \left( \frac{\mathrm{1400~MHz}}{\mathrm{325~MHz}} \right) ^{-\alpha} \, ,
\end{equation}
where $D_\mathrm{L}$ is the luminosity distance, and $\alpha$ is the synchrotron spectral index. 
Besides using as the synchrotron spectral index the calculated values of $\alpha_{\rm 1.4\, GHz}^{\rm 325\, MHz}$, we also computed the $L_{\rm 1.4\, GHz}$ values by assuming that $\alpha_{\rm 1.4\, GHz}^{\rm 325\, MHz}=0.8$. The 1.4~GHz luminosities are reported in columns~(5) and (6) in Table~\ref{tab:radio}.

\begin{figure*}[ht!]
\begin{center}
\includegraphics[bb=0 140 432 382, scale=0.53]{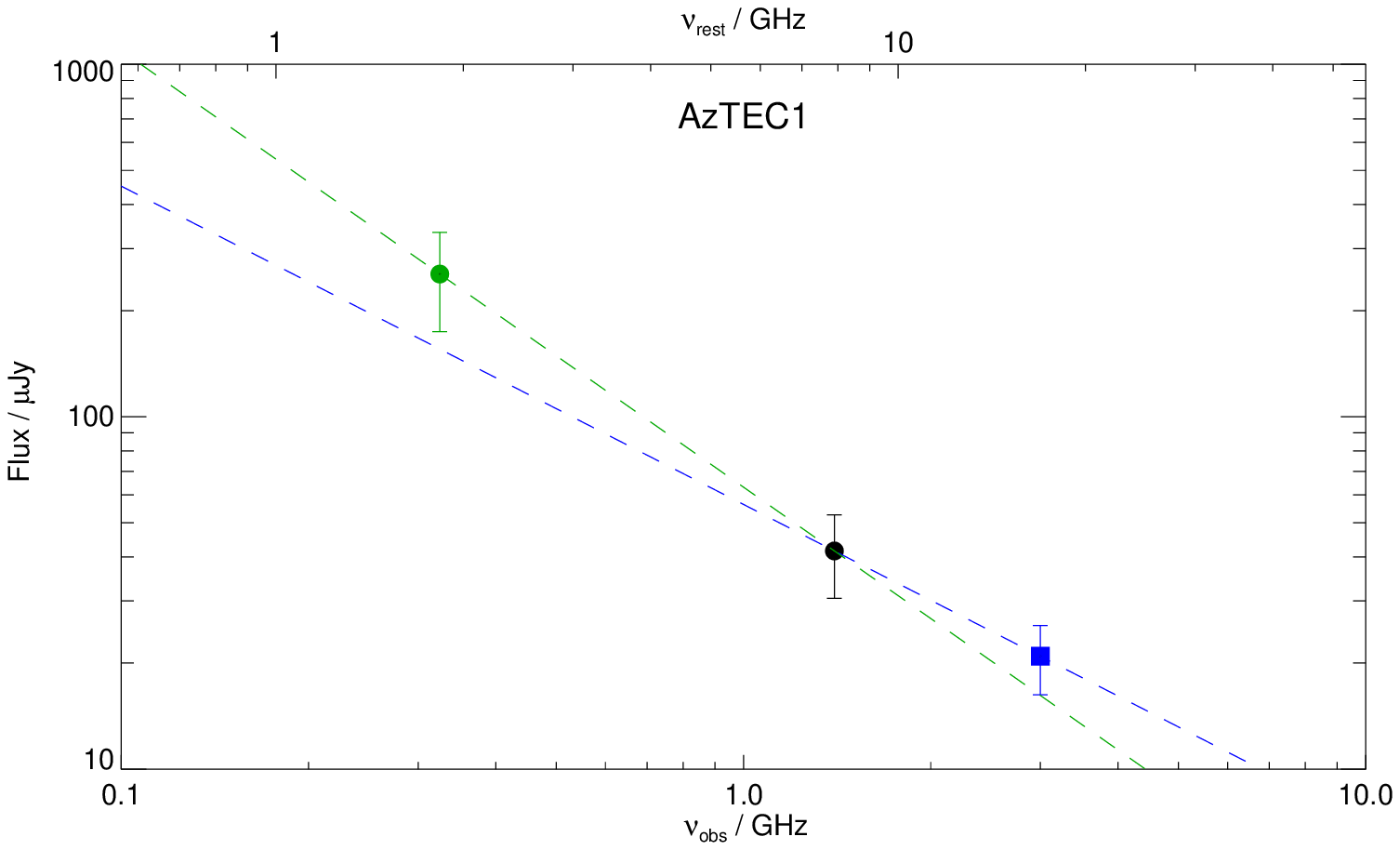}
\includegraphics[bb=0 140 432 432, scale=0.53]{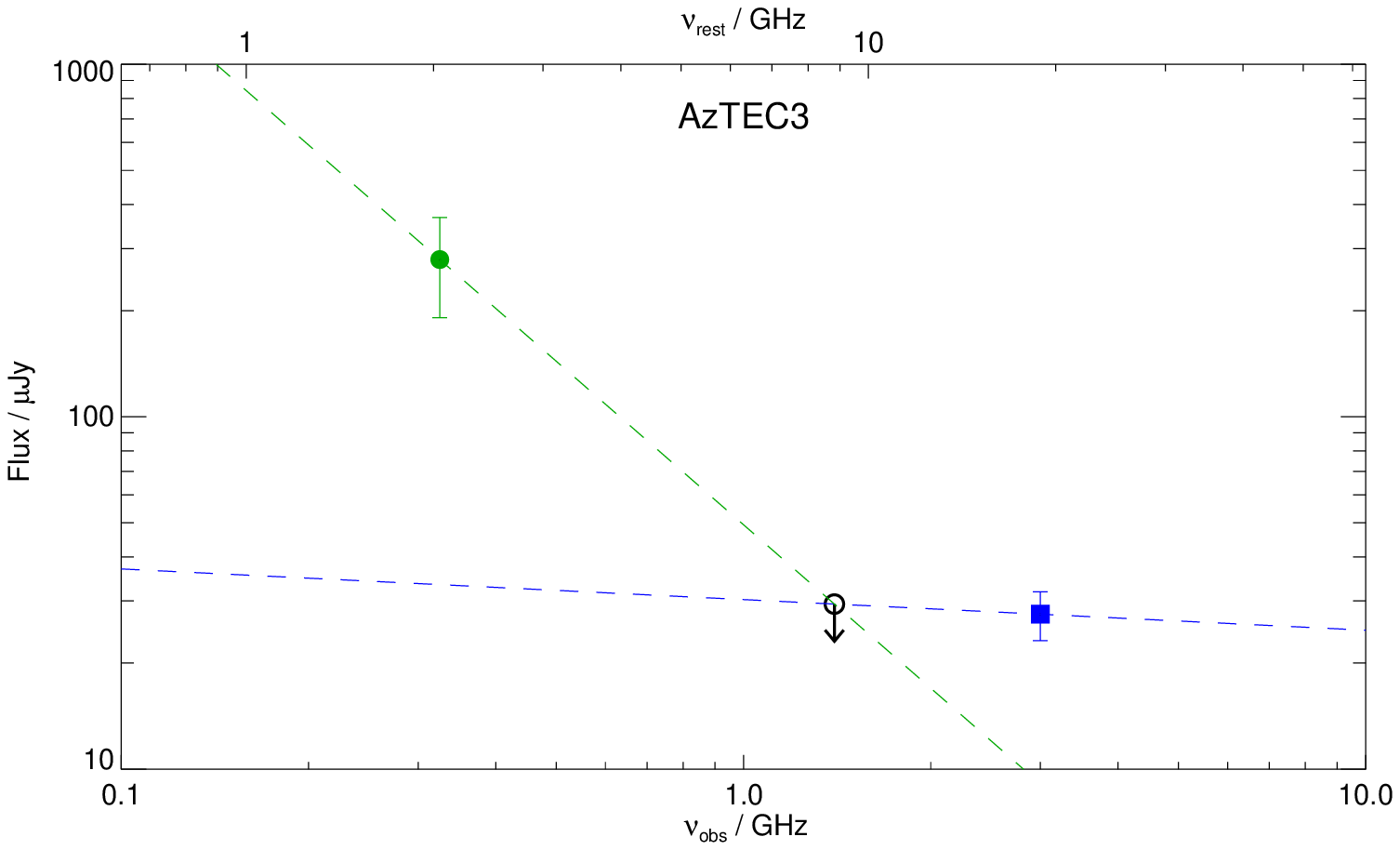}\\
\includegraphics[bb=0 140 432 432, scale=0.53]{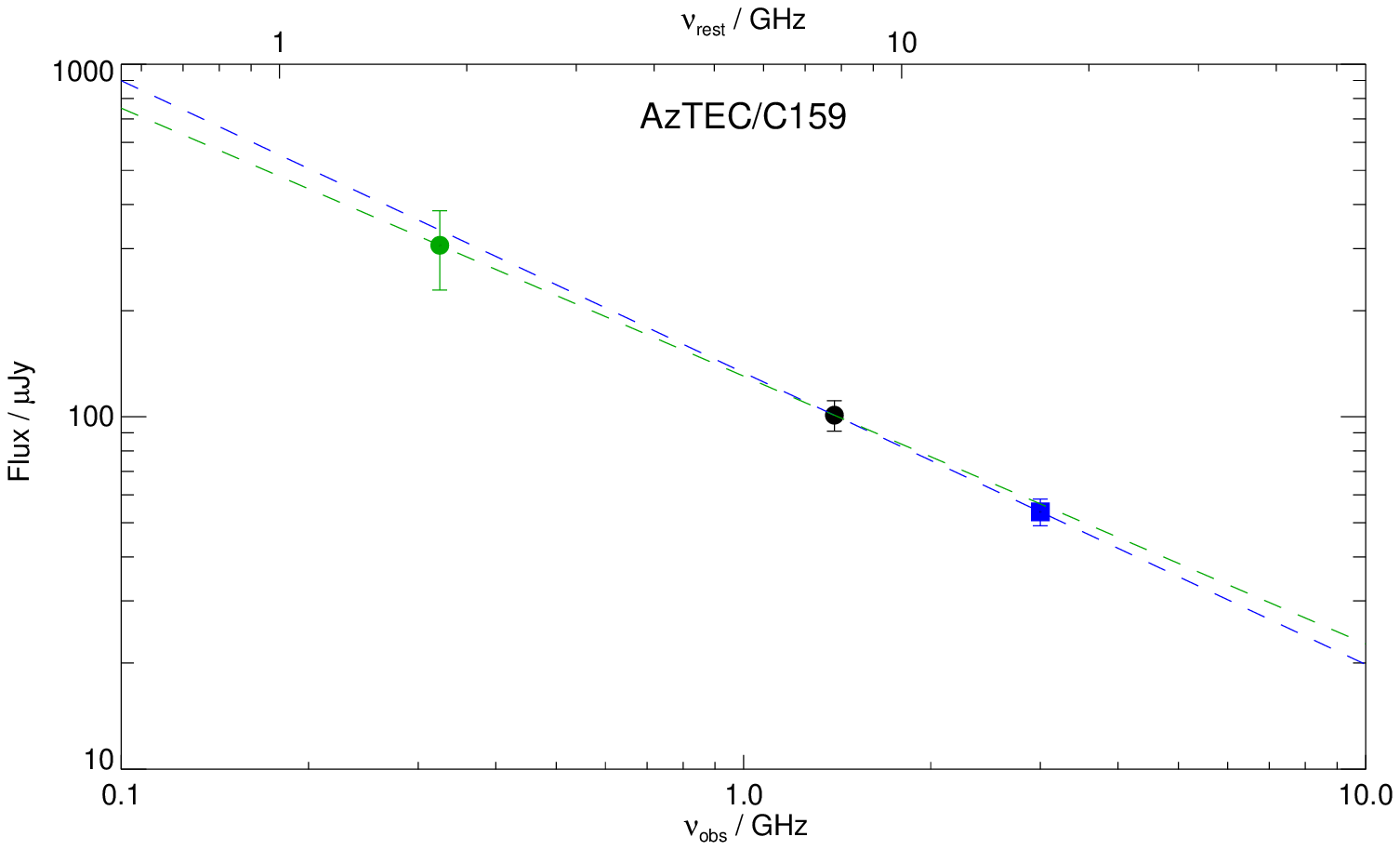}
\includegraphics[bb=0 140 432 332, scale=0.53]{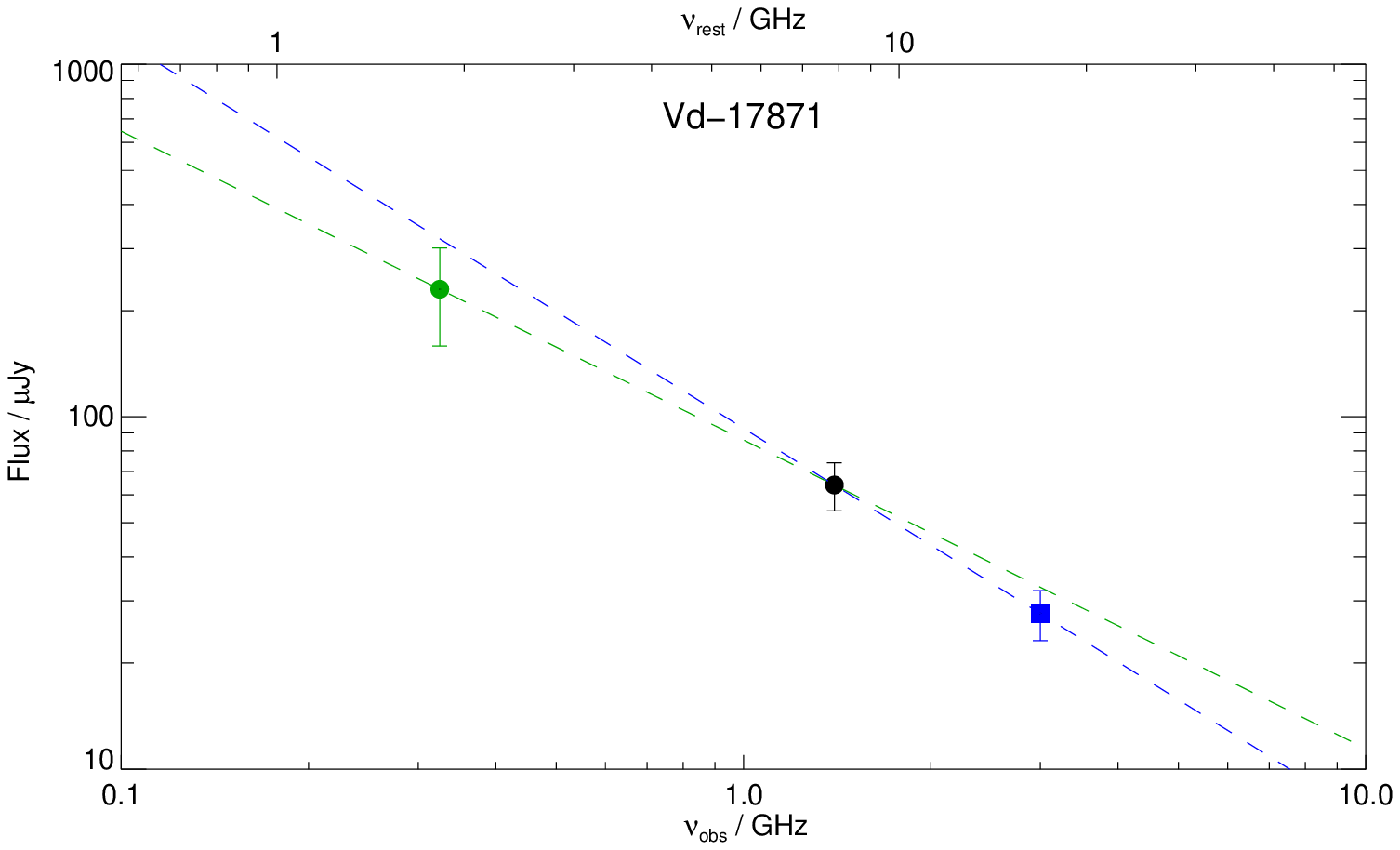}\\
\includegraphics[bb=0 20 432 432, scale=0.53]{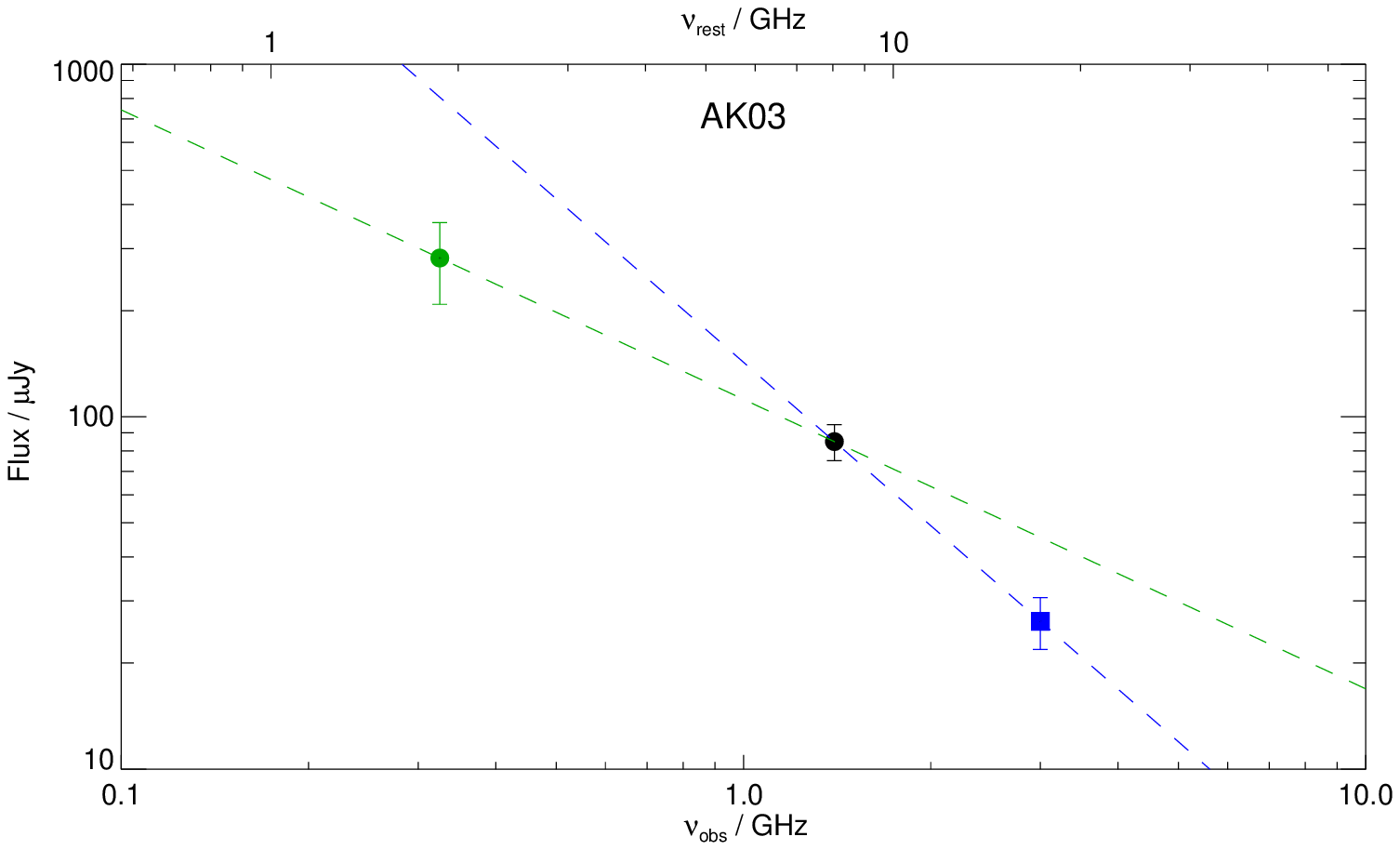}
\includegraphics[bb=0 20 432 432, scale=0.53]{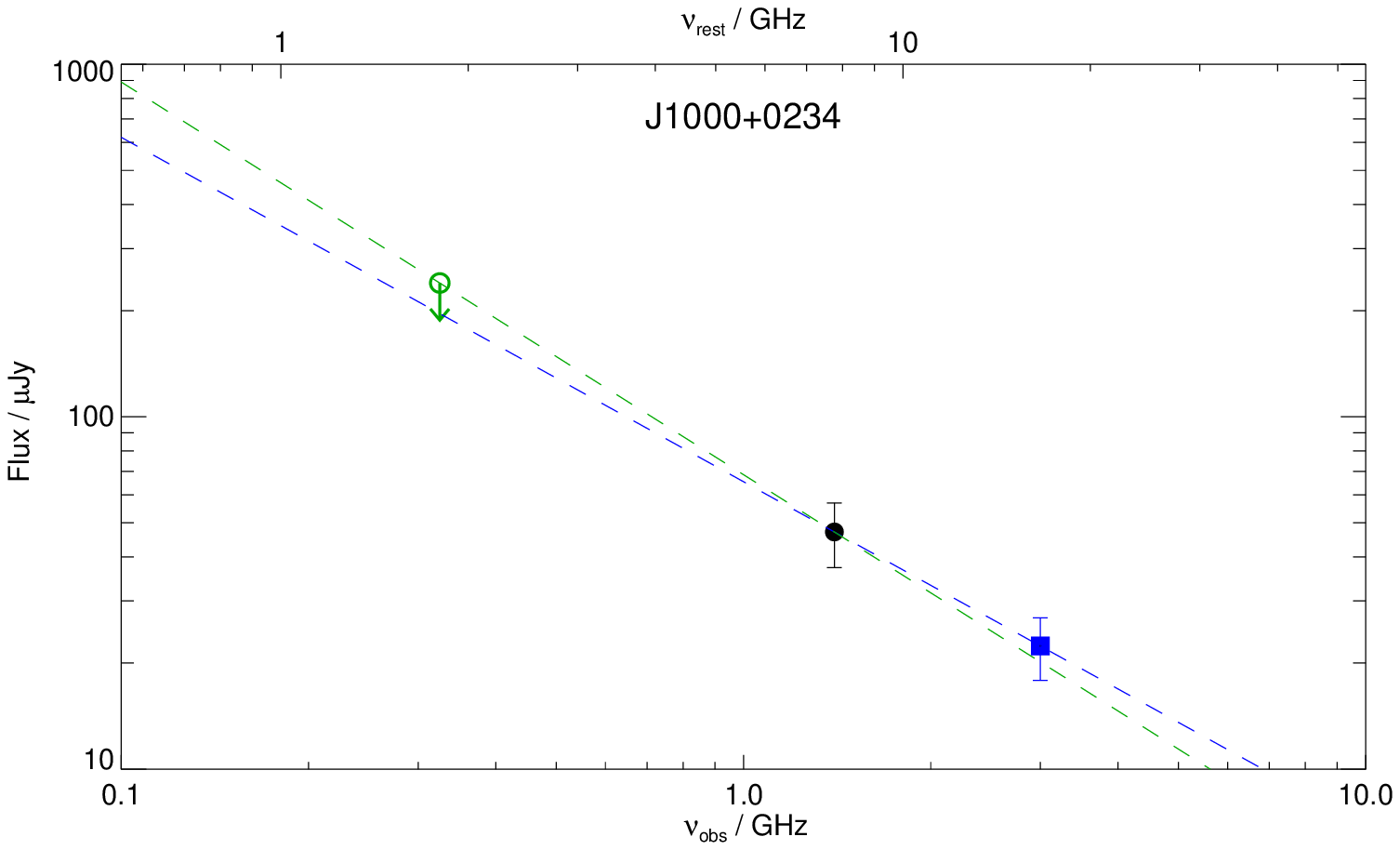}\\
\caption{The radio SEDs of the studied SMGs. The dashed lines in 
each panel represent the spectral slope taking 325~MHz (green point and error-bar) and 1.4~GHz (black  point and error-bar; green dashed line) and 1.4~GHz and 3~GHz (blue  point and error-bar) data-points (blue dashed line) into account. The resulting radio spectral indices ($\alpha$), 
here defined as flux~$\propto \nu^{-\alpha}$, are indicated in each panel. The magenta points with error bars indicate the spectral window-stacked fluxes (see text for more details). 
\label{fig:synch}}
\end{center}
\end{figure*}

\subsection{Summary of SED-derived properties of $z\gtrsim4$ SMGs in COSMOS and comparison with literature}
\label{sec:summarySmall}

As extensively studied by Dwek et al. (2011), various degeneracies in SED-derived properties exist. We also caution that a debate in the literature exists regarding the optimal model library that gives the most accurate stellar masses for SMGs. Stellar masses have been shown to systematically vary by up to $\sim0.5$~dex if different star-formation histories (\cite{michalowski2010}, 2012; \cite{hainline2011}) or IMFs (e.g., \cite{dwek2011}) are assumed, and often yield values higher than the inferred dynamical masses (\cite{michalowski2010}).  We refer the reader to Dwek et al. and Micha{\l}owski et al. for a detailed discussions on possible degeneracies and systematics, and below we summarize the results of the SED modeling performed as described above.

Our SED fitting yields that our $z>4$ SMGs contain a dominant young stellar population with ages in the range of $\sim100-700$~Myr, they already have high stellar masses (in the range of $5\times10^{10}-4\times10^{11}$~\msol), high dust luminosities ($L_\mathrm{dust}\sim10^{13}$~\lsol), and substantial dust reservoirs ($M_\mathrm{dust}\sim10^{9}$~\msol). If the gas-to-dust mass ratio is $\sim100$, the cold gas masses of our sources are comparable to their stellar masses. As expected, their SFRs are high, $\gtrsim400$~\msolyr, and scale up to a few thousand solar masses per year. 
We note that the derived stellar ages, particularly in the case of faint sources, can suffer from larger uncertainties. For comparison, Wiklind et al. (2014) recently derived a mean (median) stellar age of 0.9 (0.7) Gyr for their sample of SMGs in the CANDELS area of the GOODS-S field. The average IR luminosity we derive, based on the DL07 dust model, is $\sim1.3\times10^{13}$~\lsol, while the dust temperatures range from 39 to 48~K. 
Although large dust masses, warm dust temperatures, and high SFRs are typical for all SMGs (see \cite{casey2014} for a recent review), the derived properties of our $z>4$ SMGs put them at the high end of the distribution of SMG properties. Based on the ALESS SMG sample, Swinbank et al.\ (2014) find average dust masses of $M_\mathrm{dust}=(3.6\pm3)\times10^8$~\msol, IR luminosities of $L_\mathrm{IR}=(3.0\pm0.3)\times10^{12}$~\lsol, and SFRs of $300\pm30$~\msolyr, thus systematically lower, 
by about a factor of three, than the SMGs analyzed here. 

Within the uncertainties, our results agree with those previously reported for AzTEC1, AzTEC3, and J1000+0234. For example, \smo \ et al.\ (2011) found that AzTEC1 is a young ($\sim40-700$~Myr old) starburst galaxy with a stellar mass of $M_{\star}\sim8\times10^{10}$~\msol \ (scaled to a Chabrier IMF). They furthermore found an IR luminosity for the source of $L_\mathrm{IR}=2.9\times10^{13}$~\lsol, and a dust mass of $\sim1.5\times10^9$~\msol, in agreement with the results presented here (see Tables~5 and 6). Capak et al.\ (2011) report values of $L_\mathrm{IR}=(2.2-11)\times10^{13}$~\lsol, and SFR~$>800$~\msolyr \ for AzTEC3, consistent with our SED modeling results. For J1000+0234, Capak et al. (2008) report a starburst with a stellar mass $M_{\star}>5.5\times10^{10}$~\msol, SFR~$>550$~\msolyr, and $L_{\rm IR}=(0.5-2)\times10^{13}$~\lsol, in agreement with the results presented here. Fitting the total photometry of the source, as done here, they find a $>100$~Myr old population, consistent with our results. However, deconvolving the various source components they find a young (2-8~Myr) starburst in J1000+0234.

Toft et al. (2014) studied the SED properties of all of our sources except 
AzTEC/C159. They employed the same methods as in the present work, i.e. 
MAGPHYS and the DL07 dust model, and the derived SED parameters are mostly very 
similar to those presented here. We also note that three of our target 
sources, namely AzTEC1, AzTEC3, and J1000+0234, were part of the recent 
\textit{Herschel} study by Huang et al. (2014)\footnote{The SMG J1000+0234 is 
called Capak4.55 in Huang et al. (2014).}. They only used  
250 and 350 $\mu$m flux densities (consistent with our values within the errors). 
Our derived value of $L_{\rm IR}$ for AzTEC1 is within $2\sigma$ of that reported by Huang et al. (2014; their Table~4), while our value for AzTEC3 is in very good agreement with theirs. The value for J1000+0234 is also consistent taking into account our large errors. The dust masses of AzTEC1, AzTEC3 and J1000+0234 derived by these authors are also in very good agreement with  values inferred here. 
Our derived dust temperatures, based on the assumption of optically thin 
emission and fixed dust emissivity index ($\beta=1.5$), are in good agreement with those reported by Huang et al. (2014) in their Table~3. We caution that these assumptions were necessary simplifications to fit the full sample, and note that optical depth effects could matter (see, e.g., \cite{riechers2013}).

\section{Discussion}
\label{sec:discussion}

\subsection{Infrared-radio correlation}

 We quantify the IR-radio correlation from the derived IR and radio luminosities in the standard way, i.e. via the $q$-parameter, defined as:

\begin{equation}
q_{\rm IR}=\log{\left( \frac{L_\mathrm{IR}}{3.75\times10^{12}~\mathrm{W}}\right)} - \log\left( \frac{L_\mathrm{1.4GHz}}{\mathrm{W\, Hz^{-1}}}\right)\,.
\end{equation}

\begin{figure}[!h]
\centering
\resizebox{\hsize}{!}{\includegraphics{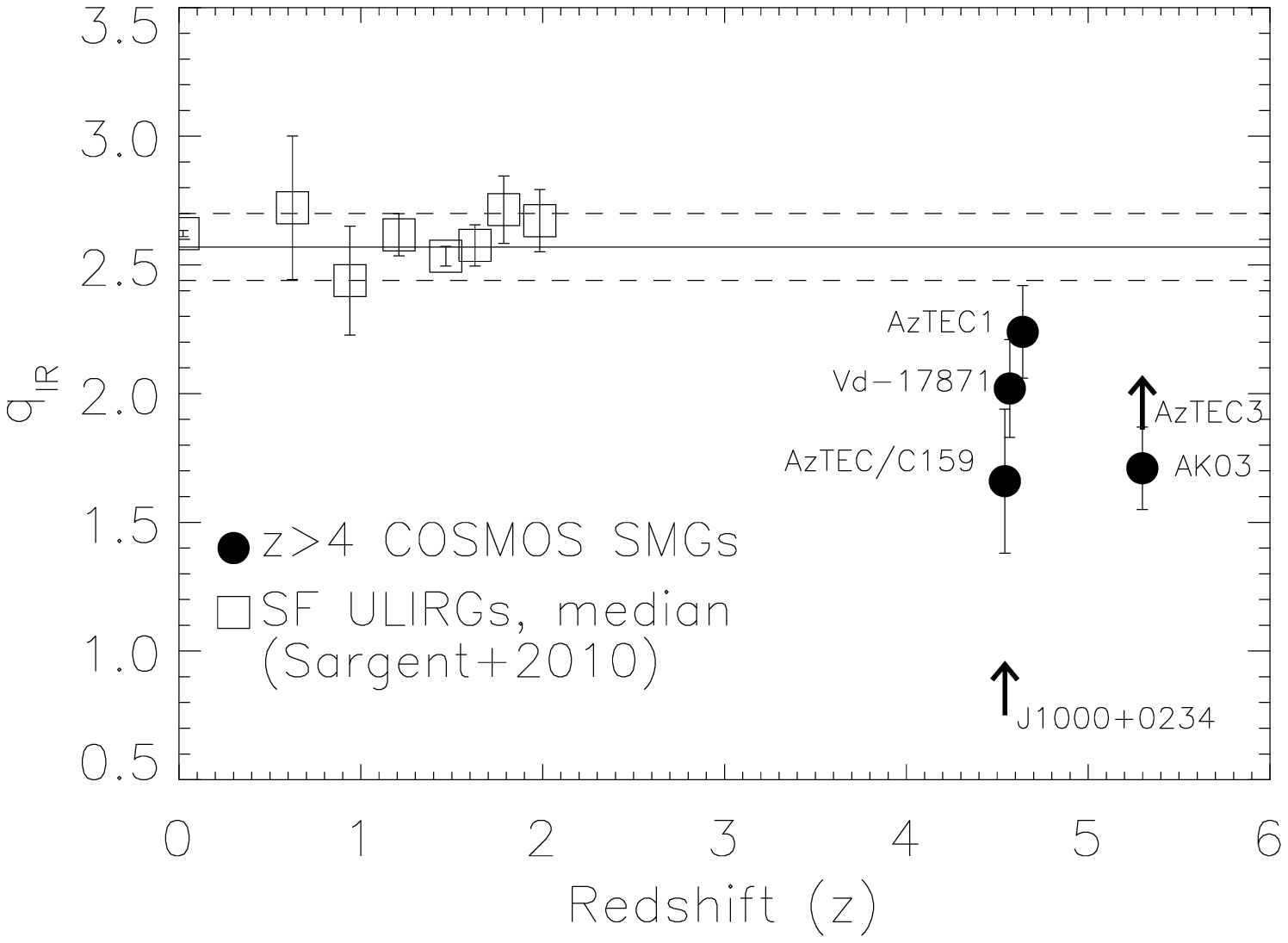}}
\caption{Infrared-radio correlation expressed via the 
$q_\mathrm{IR}$-parameter [Eq.~(2)] as a function of redshift. Shown are our $z>4$ SMGs (filled circles), and the evolution of the median of the correlation (corrected for systematic effects) for star-forming ULIRGs (open squares), adopted from Sargent et al.\ (2010a). The average $q_{\rm IR}=2.57\pm0.13$ value, found by Sargent et al.\ (2010b) for galaxies at $z<2$, is indicated by the solid line with the dashed lines denoting the errors. For AzTEC3 and J1000+0234 only lower limits to $q_{\rm IR}$ could be derived (marked with arrows).   }
\label{figure:ir-radio}
\end{figure}

We note that $L_{\rm IR} \equiv L_{\rm IR}^{\rm DL07}$ was calculated by 
integrating over the wavelength range 8--1\,000 $\mu$m. The results are shown, and compared to lower-redshift samples in Fig.~\ref{figure:ir-radio}. Although the associated errors are large (propagated from the average value of the $\pm$-error in $L_{\rm IR}^{\rm DL07}$ quoted in Table~6, and from the $\pm$-error in $L_{\rm 1.4\, GHz}$), all of our SMGs are offset from  the IR-radio relation found in the lower-redshift universe. We used the {\tt R} program package NADA (Nondetects And Data Analysis for environmental data; \cite{helsel2005}), which is an implementation of the statistical methods provided by the Astronomy Survival Analysis (ASURV; \cite{feigelson1985}) package, to take the lower limits to $q_{\rm IR}$ properly into account. This survival analysis yields a mean$\pm$standard deviation of $\langle q_\mathrm{IR}\rangle =1.95\pm0.26$ for our $z>4$ SMGs. The inferred median value is 2.02, and the 95\% confidence interval 1.75--2.16.

The inferred average $q_{\rm IR}$ value for our SMGs is lower than the $q_\mathrm{IR}$ value for lower-redshift star-forming galaxies (e.g., $2.64\pm0.02$, \cite{bell2003}; $2.57\pm0.13$, \cite{sargent2010b}).\footnote{As in the present paper, the IR luminosities in the reference studies were calculated over the wavelength range $8-1\,000$ $\mu$m.} However, our result is, within the uncertainties, consistent with the on-average lower $q_\mathrm{IR}$ values inferred for $z>4$ SMGs by Murphy (2009; $q_\mathrm{IR}=2.16\pm0.28$) and Micha{\l}owski et al.\ (2010; $q_\mathrm{IR}=2.32\pm0.20$). 

A lower $q_{\rm IR}$ value [Eq.~(2)] can arise either due to an underestimate of the IR luminosity or an overestimate of the radio luminosity. The IR luminosities (derived using the DL07 dust model) for our sources in the range of $\sim6.3\times10^{12}-2.5\times10^{13}$ L$_{\sun}$ are unlikely to be underestimated. On the other hand, our 1.4~GHz rest-frame radio luminosities have been derived using the 325~MHz observations, very close to 1.4~GHz rest-frame, thus minimizing uncertainties in the calculation (such as an assumed spectral index). This suggests that the deviation is physical. We also note that it is possible that the discrepancy arises due to the properties of our limited sample of 6 spectroscopically confirmed $z>4$ SMGs. 
Thomson et al. (2014) recently derived a median value of $q_{\rm IR}=2.56 \pm 0.05$ for their sample of 52 ALMA SMGs. The authors also found  evidence that the value of $q_{\rm IR}$ changes as a function of source evolutionary stage, and that most of the sources follow the model tracks from Bressan et al. (2002) for evolving starburst galaxies, i.e. the expected $q$ is high ($q_{\rm IR}\sim3$) in the very young starburst phase, then it decreases reaching a minimum ($q_{\rm IR}\lesssim 2$) at an age of about 40 - 50~Myr and finally increases to $q_{\rm IR}\sim2.5$ at later times. The low $q_{\rm IR}$ values we derived would then support the idea that the sources are in a relatively  early stage of evolution.

\subsection{Morphology and sizes of star-formation regions} 

As evident from the \textit{HST}, and stacked UltraVISTA stamps of our SMGs, shown in Fig.~\ref{fig:stamps}, AzTEC1 and AzTEC3 appear compact, while J1000+0234, Vd-17871, and AK03 show signs of clumpiness and  multiple components. AzTEC/C159 is a faint source undetected in UltraVISTA $YJHK_{\rm s}$, but detected in the radio at high significance with clearly resolved 3~GHz emission. 

Capak et al.\ (2008) have demonstrated that all J1000+0234 (compact and extended) components are at the same redshift given that Ly$\alpha$ emission was observed for all of them, and suggested that the components were merging. CO observations of J1000+0234 independently suggest the source is in the process of merging (\cite{schinnerer2008}). 
Karim et al.\ (in prep.) have shown that the two components of Vd-17871 lie at comparable spectroscopic redshifts, implying that they are associated and perhaps merging. Based on  photometric redshift the same is inferred for AK03-S (Section~\ref{sec:sample}). Thus, our sample consists of approximately equal number of compact and clumpy systems, consistent with the idea that SMGs form a morphologically heterogenous sample (e.g., \cite{hayward2012}). It is also possible to see mergers at different stages, for example before or after the coalescence process when regular disc-like geometry is not expected (cf.~\cite{targett2011}, 2013). For comparison, Tasca et al. (2014) found that about 20\% of star-forming galaxies with stellar masses of a few times $10^{10}$ M$_{\sun}$ are involved in major mergers. Based on ALMA data Riechers et al. (2014) find that AzTEC3 is likely to be a merger.

We have made use of our 3~GHz VLA-COSMOS observations at a resolution of $0\farcs6 \times 0\farcs7$ to infer the sizes of the star-forming regions in our SMGs (assuming that the 3~GHz emission is due to star-formation processes). As evident from \f{figure:3GHz} \ none of the sources shows several radio-emitting components. We find  AzTEC/C159 to be clearly resolved at 3~GHz and estimate a deconvolved size of the source (FWHM) at 3~GHz of $0\farcs94 \times 0\farcs46$ or $6.2\times3.0$~kpc. We find  that J1000+0234 and AK03 may be marginally resolved, and their deconvolved sizes (FWHM) are $0\farcs96 \times <0\farcs7$ and $0\farcs60 \times 0\farcs41$, respectively, corresponding to $6.3\times4.6$~kpc (J1000+0234) and $3.9\times2.6$~kpc (AK03). The remaining sources are unresolved at a resolution of $0\farcs7$ setting upper limits on their star-formation region sizes (FWHM) to $\leq4.6$~kpc (AzTEC1), $\leq4.3$~kpc (AzTEC3), and $\leq4.6$~kpc (Vd-17871). Note that high-resolution SMA 890~$\mu$m continuum and ALMA $[\ion{C}{II}]$ measurements of AzTEC1 and AzTEC3, respectively, set an upper limit of the star formation region size of FWHM~$\leq1.3$~kpc for AzTEC~1, and  $3.9$~kpc for AzTEC3 (\cite{younger2008}; \cite{riechers2010}, 2014). Based on these radio and (sub-)mm measurements, we estimate the average size of the star-formation regions of our $z>4$ SMGs using the ASURV survival analysis. This yields a mean major axis of $0\farcs63 \pm 0\farcs12$ and minor axis of $0\farcs35 \pm 0\farcs05$ for our sources, and it roughly corresponds to a size of $4.1\times2.3$~kpc$^2$ (assuming $z=4.5$). Considering J1000+0234 and AK03 unresolved, the mean major axis decreases to $0\farcs43 \pm 0\farcs14$. We  note that the NIR effective radii derived by Toft et al. (2014) for AzTEC1 and -3 ($<2.6$ and $<2.4$ kpc), J1000+0234 ($3.7\pm0.2$ kpc), and AK03 ($1.6\pm0.6$ kpc) agree well with the above radio FWHM sizes. 
The sizes inferred here are in agreement with the sizes inferred for lower-redshift SMGs, but larger than the sizes of local ULIRGs (e.g. \cite{chapman2004}; \cite{tacconi2006}; \cite{rujopakarn2011}; \cite{simpson14b}). In their recent ALMA band-7 continuum follow-up of a sample of Scuba-2 detected SMGs in the UDS
field, \cite{simpson14b} determine an average extent of FWHM$\sim$0.3". While they report
average 1.4 GHz radio continuum sizes that are about twice as large, the cold dust emitting region probed
by their 870$\mu$m continuum observations hence appears to be consistent with the lower bound we derive
on average from our high resolution JVLA radio data. Whether the average size of the dust emitting region of
our specifically $z>4$ selected sources is significantly smaller as found for the Simpson et al. (2014) SMG sample remains
to be clarified based on adequately high resolution interferometric follow-up.  

\subsection{Relationship between the IR luminosity and dust temperature}

\begin{figure}
\begin{center}
	\includegraphics[bb=0 0 504 360, width=\columnwidth]{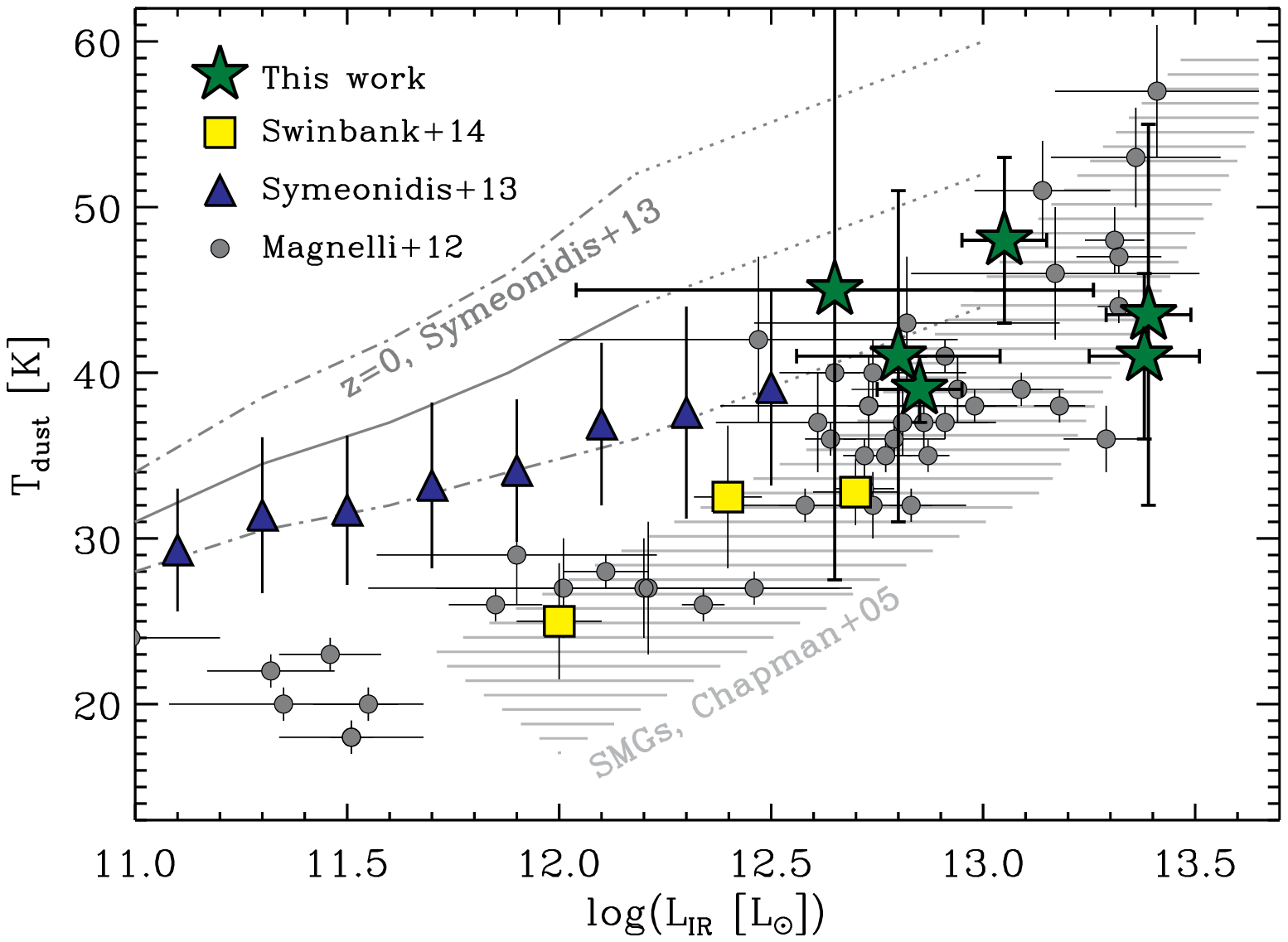}
	\caption{The dust temperature plotted as a function of IR luminosity. The values obtained for our SMGs are shown with green star symbols. The other symbols and lines mark values and correlations from the quoted reference studies (see text for details). The shaded region illustrates the correlation found by Chapman et al. (2005) for their sample of SMGs. 
	\label{fig:lt}
	}
\end{center}

\end{figure}

Previous studies of populations of infrared-luminous galaxies out to
high redshift (including SMGs) have found a relation between dust
temperature ($T_\mathrm{dust}$) and IR luminosity ($L_\mathrm{IR}$),
such that, on average, $T_\mathrm{dust}$ increases with
$L_\mathrm{IR}$ (e.g., \cite{chapman2005}; \cite{magnelli2012};
\cite{symeonidis2013}; see also \cite{swinbank2014}). The scatter
in this relation strongly depends on the wavelength selection (e.g.,
\cite{magnelli2012}), but there appears to be a trend for SMGs to
have higher $L_\mathrm{IR}$ than nearby/low-$z$ ULIRGs at the same
$T_\mathrm{dust}$ (expressed by some authors as a lower $T_\mathrm{dust}$
at a given $L_\mathrm{IR}$).\footnote{Note that these populations only
overlap at the lower-$L_\mathrm{IR}$ end of SMGs.} Assuming that the same
$T_\mathrm{dust}$ can only be maintained by the same IR luminosity
surface densities ($\Sigma_\mathrm{IR}$), this trend likely indicates more
spatially-extended IR-emitting regions in SMGs, at a given
$T_\mathrm{dust}$. This picture is
consistent with the large, typically 2--4\,kpc sizes of SMGs observed
at submillimeter wavelengths (e.g., \cite{tacconi2008}), which
indicate wide-spread star formation fueled by large quantities of cold
gas (e.g., \cite{ivison2011}; \cite{riechers2011}). Like other
star-forming galaxy populations at similar redshifts (e.g., \cite{daddi2008}, 2010; \cite{tacconi2010}, 2013), 
SMGs also appear to have higher gas mass fractions than nearby starbursts and ULIRGs 
(e.g., \cite{riechers2011}; \cite{ivison2011}), and thus, higher gas and
dust masses than nearby galaxies at the same $L_\mathrm{IR}$. This is consistent with the interpretation of the
$T_\mathrm{dust}$--$L_\mathrm{IR}$ relation by Symeonidis et al.\
(2013), but we note that the relation to the dust mass is likely
indirect at best, since the dust in SMGs is likely optically thick at
the peak wavelengths that are most decisive for the estimates of
$T_\mathrm{dust}$ and $L_\mathrm{IR}$.

To put the derived properties of our $z>4$ SMGs in context of the
general population of IR galaxies, in \f{fig:lt} \ we show dust temperature
versus IR luminosity for our SMGs, and compare this to various IR- and
submm-selected samples presented in the literature: the locus of local
($z\lesssim0.1$) \textit{IRAS}-selected IR galaxies
(\cite{symeonidis2013} and references therein),
\textit{Herschel}-selected $0.1<z<2$ IR galaxies
(\cite{symeonidis2013}), a compilation of SMGs from various fields
(\cite{magnelli2012}), the SMG sample from Chapman et al.\ (2005), and
the ALESS SMG sample (\cite{swinbank2014}).

Our $z>4$ SMGs fall in the scatter of the $z\sim2$ SMG sample from
Chapman et al.\ (2005) (albeit towards the high-$L_\mathrm{IR}$ end
due to their initial selection), consistent with their measured, few
kpc sizes at radio and/or submillimeter wavelengths. Some of the $z>4$
SMGs fall towards the high-$T_\mathrm{dust}$ end, consistent with the
different K-correction at higher redshifts in the (sub)millimeter
bands (e.g., \cite{chapman2005} -- maybe also add Blain et
al. 2004-ish). These systems consistently show high IR
luminosity and SFR surface densitites 
(e.g., \cite{younger2008}; \cite{riechers2014}), approaching those of so-called
``maximum starbursts'' (e.g., \cite{scoville2003}; \cite{thompson2005}). 
Our inferred, highest dust masses are also consistent with this picture.

\section{Summary}
\label{sec:summary}

To constrain the physical characteristics of high-redshift SMGs, we have 
carried out a study of a sample of six SMGs in the COSMOS field with 
spectroscopically confirmed redshifts $z>4$ (the average $z$ being 4.7). 
Our main findings are summarized as follows:

\begin{itemize}

\item[$\bullet$] Based on new, VLA-COSMOS 3~GHz radio data at resolution of $0\farcs6\times0\farcs7$ we estimate the sizes of the star-formation regions of our $z>4$ SMGs. Using additional millimetric imaging from literature for AzTEC1 and AzTEC3 and taking limits into account using the ASURV package we estimate a mean size of $( 0\farcs63 \pm 0\farcs12 ) \times (0\farcs35 \pm 0\farcs05 )$ or about $4.1 \times 2.3$ kpc$^2$ (assuming $z=4.5$) for our $z>4$ SMGs (or lower if we take the two possibly resolved SMGs, J1000+0234 and AK03, as unresolved).  These values are consistent with those derived for lower-redshift SMGs, yet higher than for local ULIRGs. \\ 

\item[$\bullet$] The SED analysis performed with the MAGPHYS code suggests 
that the stellar masses in our SMGs are in the range 
$M_{\star}\sim0.5-3.9\times10^{11}$ M$_{\sun}$. The resulting dust masses and 
luminosities are found to be $\sim3.6\times10^8-5.6\times10^{10}$ M$_{\sun}$ 
and $\sim0.14-3.4\times10^{13}$ L$_{\sun}$. The inferred ages of the oldest 
stars in these systems range from 
about 100 to $\sim700$ Myr, with an average age of $\sim280$ Myr, and with 5 out of 6 SMGs having estimated ages smaller than $\sim200$~Myr. The dust 
properties were also determined through fitting the SEDs with modified 
blackbody curves and employing the Draine \& Li (2007) dust model. 
The former yields the values $M_{\rm dust}\sim0.3-1.3\times10^9$ M$_{\sun}$ and 
$L_{\rm IR}\sim 4\times10^{11}-2.3\times10^{13}$ L$_{\sun}$, while the latter 
method gives $M_{\rm dust}\sim1-5\times10^9$ M$_{\sun}$ and 
$L_{\rm IR}\sim0.5-2.5\times10^{13}$ L$_{\sun}$. The derived dust temperatures, 
$\sim39-48$ K, are relatively warm compared to lower-redshift, but lower 
$L_{\rm IR}$ SMGs. Finally, the star formation rates we determine, 
$\sim450-2\,500$ M$_{\sun}$~yr$^{-1}$, conform with the idea that the sources 
are very dusty starburst galaxies. \\ 

\item[$\bullet$] Using new 325~MHz (GMRT), 1.4~GHz (VLA), and 3~GHz (VLA) data 
we  derive the 1.4~GHz luminosities for our SMGs, found to be in the 
range of about $10^{25}$~W~Hz$^{-1}$ up to possibly 
$2.4\times10^{25}$~W~Hz$^{-1}$ (when using spectral indices drawn from 0.325 
and 1.4 GHz fluxes). \\

\item[$\bullet$] Using the derived (\cite{draine2007}) IR and 1.4~GHz 
luminosities we investigate the IR-radio correlation for our SMGs, as 
quantified by the $q$ parameter. We find it to be lower for our $z>4$ 
SMGs ($\langle q \rangle=1.95\pm0.26$) compared to that found for 
lower-redshift galaxies. This may be due to selection effects and supports the idea that the sources 
are in a relatively early stage of evolution. \\ 

\item[$\bullet$] Inspection of the relationship between IR luminosity and dust 
temperature shows that our SMGs lie at the high end of the correlation, consistent with an extrapolation of the high-redshift  $L_{\rm IR}-T_{\rm dust}$ relation previously inferred for \textit{Herschel}-selected $0.1<z<2$ galaxies. Thus, we conclude that at fixed $L_{\rm IR}$ the cold dust $L_{\rm IR}-T_{\rm dust}$ of high-redshift SMGs is likely due to more extended star-formation regions in high-redshift SMGs relative to local ULIRGs and higher dust masses in high-redshift galaxies, compared to lower-redshift SMG samples, as has been previously suggested.\\

\end{itemize}

In summary, we find that our sample consists of a fair 
mix of compact and clumpy systems, consistent with the idea that (sub-)mm 
galaxy identification may select physically heterogeneous samples of galaxies. 
The compact appearance might be an indication of a merger remnant rather than 
a regular star-forming disk galaxy. Thus, further studies are required to 
examine how $z>4$ SMGs are related to  $z<3$ SMGs, and other populations, such 
as massive ellipticals at $z\sim2$ and locally, as well as to quantify the 
role played by  SMGs in galaxy formation and evolution.

\begin{table*}
\caption{The sample of spectroscopically confirmed $z>4$ SMGs in the COSMOS 
field.}
\begin{minipage}{2\columnwidth}
\small
\centering
\renewcommand{\footnoterule}{}
\label{table:sample}
\begin{tabular}{c c c c c c c}
\hline\hline 
Source & $\alpha_{2000.0}$ & $\delta_{2000.0}$ & $z_\mathrm{spec}$ & $S_{\lambda}$ & $\lambda$ & Instruments \\
       & [\degr] & [\degr] & & [mJy] & [mm] & \\
\hline
AzTEC1 & 149.92859 & 2.49394 & 4.640 &  $9.3 \pm 1.3 $ & 1.1& JCMT/AzTEC, SMA, Keck~II/DEIMOS \\ 
AzTEC3 &  150.08624 & 2.58903 & 5.298 &  $5.9 \pm 1.3 $ & 1.1& JCMT/AzTEC, SMA, Keck~II/DEIMOS \\ 
AzTEC/C159 & 149.87671 & 1.92432 & 4.569 & $3.3\pm1.3$ & 1.1 & ASTE/AzTEC,VLA, Keck~II/DEIMOS \\
J1000+0234 &  150.22717 & 2.57644 & 4.542 &  $3.4 \pm 0.7 $ & 1.2 & IRAM-30m/MAMBO, PdBI, Keck~II/DEIMOS \\ 
Vd-17871 &  150.36284 & 2.14883   & 4.622 &  $2.5 \pm 0.5 $ & 1.2 & IRAM-30m/MAMBO, PdBI, Keck~II/DEIMOS, VLT/VIMOS  \\ 
AK03 & 150.07841 & 2.47061  & 4.747 &  $11.42\pm4.98 $ & 0.450 & JCMT/SCUBA2, VLA, Keck~II/DEIMOS \\ 
\hline 
\end{tabular} 
\tablefoot{Column~(6) gives the (sub)mm-wavelength used for initial identification of the source as an SMG, and the last column lists the instruments used for initial (sub)mm-detection, and counterpart and redshift association.}
\end{minipage}
\end{table*}

\begin{table*}
\caption{\textit{Herschel} PACS/SPIRE photometry for our $z>4$ SMG sample. \label{tab:photherschel}}
\renewcommand{\footnoterule}{}
\label{tab:herschelphot}
\begin{tabular}{lcccccccccccccccccccc}
\hline\hline
Source & $S_{100\mathrm{\mu m}}$ & $S_{160\mathrm{\mu m}}$ & 
    $S_{250\mathrm{\mu m}}$ & $S_{350\mathrm{\mu m}}$ & $S_{500\mathrm{\mu m}}$  & Flag\tablefootmark{*} \\
  & [mJy] & [mJy]& [mJy] & [mJy] & [mJy]  & \\
\hline
AzTEC1 & $<6.8$ & $<13.6$ & $19.5\pm6.0$  & $29.8 \pm 7.6$ & $28.8 \pm 9.0$ & 1\\ 
AzTEC3 & $<6.8$ & $<13.6$ & $22.3 \pm 7.1$ & $21.1 \pm 7$ & $<32.0$ & 1 \\ 
 AzTEC/C159   &        $<6.8$   &   $<13.6$   &  $6.2\pm1.9$  &     $<10.8$     &   $<20.4$   &  2 \\
J1000+0234 & $<6.8$ & $<13.6$ & $<15$& $<20.0$ & $<30.0$  & 3\\ 
 Vd-17871 & $<6.8$ & $<12.0$ & $9.5 \pm 2.0$ & $12.3 \pm 2.$4 & $13.0 \pm 3.7$ & 1 \\ 
AK03 & $<6.8$ & $<13.6$ & $<8.0$ & $<12.0$ & $<20.0$  & 1\\ 
\hline
\end{tabular}
\tablefoot{\tablefoottext{*}{Flag=1: Secure FIR detections; Flag=2: FIR detections are confused but stable; Flag=3: FIR detections are strongly affected by confusion. The quoted upper limits are $4\sigma$.}}
\\
\end{table*}

\begin{table*}
\caption{Sub-mm to radio photometry for our $z>4$ SMG sample. \label{tab:photmm}}
\label{tab:aztecphot}
\begin{tabular}{lcccccccccccccccccccc}
\hline\hline
Source & $S_{870\mathrm{\mu m}}$ & $S_{890\mathrm{\mu m}}$ & 
    $S_{1.1\mathrm{mm}}$ & $S_{1.2\mathrm{mm}}$ & 
    $S_\mathrm{10cm}$ & $S_\mathrm{20cm}$ & $S_\mathrm{90cm}$ \\
 & [mJy] & [mJy]  & 
    [mJy]  & [mJy] & [mJy]   & 
    [$\mu$Jy]  & [$\mu$Jy] \\

\hline
AzTEC1 & $12.6 \pm 3.6$ & $15.6 \pm 1.1$ & $9.3 \pm 1.3$ & -- &  $20.90 \pm 4.64$ & $41.6 \pm11.1$ & $254.0 \pm   79.7$ \\ 
AzTEC3 & -- & $8.7 \pm 1.5$ & $5.9 \pm 1.3$ & -- &   $ 27.51\pm  4.36$   & $<29.4$\tablefootmark{a} & $279.3 \pm   88.2$\\
AzTEC/C159 & -- & -- & $3.3\pm1.3$ & -- &  $53.73\pm 4.64$ & $101.0\pm 10.0$ & $306.5\pm   77.8$\\
J1000+0234 & -- & -- & $3.4\pm0.7$ & -- &     $22.37 \pm 4.51$  & $47.1\pm9.8$  & $<239.6$\tablefootmark{a}\\ 
Vd-17871 & -- & -- & $4.2 \pm 1.3$ & $2.5 \pm 0.5$ &   $27.64 \pm  4.48$   & $64.0\pm10.0$ & $230.0\pm  71.4$\\ 
AK03 & -- & -- & -- & -- &  $26.27 \pm  4.40$ & $85.0\pm10.0$ & $282.0\pm 73.61$\\ 
\hline
\end{tabular}
\tablefoot{
\tablefoottext{a}{$3\sigma$ upper limit.} 
}
\end{table*}

\begin{table*}
\caption{Radio properties of our $z>4$ SMG sample.  }
\begin{minipage}{2\columnwidth}
\small
\centering
\renewcommand{\footnoterule}{}
\label{tab:radio}
\begin{tabular}{l c c c c c c c}
\hline\hline 
Source & size\tablefootmark{a} & $\alpha^\mathrm{325MHz}_\mathrm{1.4GHz}$ & $\alpha^\mathrm{1.4GHz}_\mathrm{3GHz}$ & $L_{\rm 1.4\,GHz}$ & $L_{\rm 1.4\,GHz}$\tablefootmark{b} & $q_{\rm IR}$ \\
       & [$\arcsec$] & &  &  [W~Hz$^{-1}$] & [W~Hz$^{-1}$] &  \\
\hline
AzTEC1 &$<0.7$ & $1.24\pm 0.28$ &  $0.90\pm0.46$  & $1.4\pm0.4\times10^{25}$ & $1.2\pm0.4\times10^{25}$ & $2.24\pm0.18$ \\ 
AzTEC3 & $<0.7$ & $>1.54$ & $<0.09$ & $<3.1\times10^{25}$ & $1.8\pm0.6\times10^{25}$ & $>1.86$ \\ 
AzTEC/C159 &$0.94\times0.46$ & $0.76\pm0.19$ & $0.83\pm0.17$ & $1.4\pm0.4\times10^{25}$ & $1.4\pm0.4\times10^{25}$ & $1.71\pm0.16$ \\
J1000+0234 & $0.96 \times <0.7$\tablefootmark{c} & $<1.11$ & $0.98\pm0.38$ & $<1.2\times10^{25}$ & $<1.1\times10^{25}$ & $>0.75$ \\ 
Vd-17871 & $<0.7$  & $0.88\pm0.24$ & $1.1\pm0.3$ & $1.1\pm0.4\times10^{25}$ & $1.1\pm0.3\times10^{25}$ & $2.02\pm0.19$ \\
AK03 & $0.60\times0.41$\tablefootmark{c} & $0.82\pm0.20$ & $1.54\pm0.27$ & $1.4\pm0.4\times10^{25}$ & $1.4\pm0.4\times10^{25}$ & $1.66\pm0.28$ \\ 
\hline 
\end{tabular} 
\tablefoot{
\tablefoottext{a}{Deconvolved major and minor axes or upper limit corresponding to the $0\farcs6 \times0\farcs7$ synthesized beam at 3~GHz.} 
\tablefoottext{b}{Calculated by assuming $\alpha^\mathrm{325MHz}_\mathrm{1.4GHz}=0.8$.} 
\tablefoottext{c}{Possibly marginally resolved, but also consistent with being unresolved.} 

}
\end{minipage}
\end{table*}

\begin{table*}
\caption{Physical properties of our $z>4$ SMG sample derived from the UV-mm SED (using MAGPHYS).  }
\renewcommand{\footnoterule}{}
\label{tab:magphys}
\begin{tabular}{lccccccccccccc}
\hline\hline
Name & $t_\mathrm{form}$ & $\tau_\mathrm{V}$ & $M_{\star}$ & $L_\mathrm{dust}$ & $M_\mathrm{dust}$ \\
& [Myr] & [mag] & [M$_\odot$] & [L$_\odot$] & [M$_\odot$]  \\
\hline
AzTEC1 & 199 & 2.897 & $7.4\times10^{10}$  & $2.2\times10^{13}$ & $3.9\times10^{9}$ \\ 
AzTEC3 & 140 & 3.952 & $1.4\times10^{11}$  & $1.1\times10^{13}$ &  $5.2\times10^{9}$ \\
AzTEC/C159 & 114 & 3.568 & $1.1\times10^{11}$  & $1.1\times10^{13}$  &  $1.6\times10^{9}$  \\ 
J1000+0234 & 149 & 31.644 & $8.7\times10^{10}$ & $7.0\times10^{12}$  & $9.1\times10^{8}$ \\ 
Vd-17871 & 199 & 2.897 & $4.6\times10^{10}$  & $1.4\times10^{13}$ & $1.5\times10^{9}$ \\
AK03-N\tablefootmark{*} & 474 & 4.721 & $1.0\times10^{11}$  &  $1.4\times10^{12}$ &  $3.6\times10^{8}$ \\
AK03-S\tablefootmark{*} & 713 & 6.243 & $3.9\times10^{11}$  &  $3.4\times10^{13}$ &  $5.6\times10^{10}$ \\
\hline
\end{tabular}\\
\tablefoot{\tablefoottext{*}{Fitted using MAGPHYS spectral template library optimized for normal galaxies.} }
\end{table*}

\begin{table*}
\caption{Physical properties derived from the IR SED of our $z>4$ SMG sample.   }
\renewcommand{\footnoterule}{}
\label{tab:ir}
\begin{tabular}{lcccccccccccccccccccc}
\hline\hline
Name & $T_\mathrm{dust}$ & $M_\mathrm{dust}$ & $M_\mathrm{dust}$ & $L_\mathrm{IR}$ & $L_\mathrm{IR}$ & SFR\tablefootmark{a}\\ 
 &  [K] &  [BB,$10^8$~M$_\odot$] &  [DL07,$10^8$~M$_\odot$] &  [BB,$10^{12}$~L$_\odot$] &  [DL07,$10^{12}~$L$_\odot$] & [DL07,M$_{\sun}$~yr$^{-1}$] \\ 
\hline \\
AzTEC1 &   41.0 $\pm$    5.0 & $  12.6^{+    3.3}_{-   2.6}$ & $  50.1^{+   13.0}_{-  10.3}$ & $  22.9^{+    8.7}_{-   6.3}$ & $  24.0^{+    8.4}_{-   6.2}$ & $2398^{+  837}_{-
 620}$ \\ [1ex]
AzTEC3 &   43.5 $\pm$   11.5 & $   7.9^{+    7.9}_{-   4.0}$ & $  25.1^{+    6.5}_{-   5.2}$ & $  15.8^{+   17.3}_{-   8.3}$ & $  24.5^{+    6.4}_{-   5.0}$ & $2454^{+  635}_{-
 504}$ \\ [1ex]
AzTEC/C159 &   39.0 $\pm$    2.0 & $   5.0^{+    1.3}_{-   1.0}$ & $  20.0^{+   30.2}_{-  12.0}$ & $   6.8^{+    1.4}_{-   1.1}$ & $   7.1^{+    1.8}_{-   1.5}$ & $ 707^{+  183
}_{- 145}$ \\ [1ex]
J1000+0234\tablefootmark{b}\ &   45.0 $\pm$   17.5 & $   5.0^{+   74.4}_{-   4.7}$ & $  39.8^{+   86.1}_{-  27.2}$ &  $14.1^{+0.0}_{-14}$ & $   4.5^{+   13.7}_{-   3.4}$ & $ 446^{+ 1373
}_{- 337}$ \\ [1ex]
Vd-17871 &   48.0 $\pm$    5.0 & $   2.5^{+    1.5}_{-   0.9}$ & $  12.6^{+    3.3}_{-   2.6}$ & $   9.8^{+    1.2}_{-   1.1}$ & $  11.2^{+    2.9}_{-   2.3}$ & $1122^{+  290
}_{- 230}$ \\ [1ex]
AK03 &   41.0 $\pm$   10.0 & $   5.0^{+    5.0}_{-   2.5}$ & $  20.0^{+   11.7}_{-   7.4}$ & $   2.0^{+    7.3}_{-   1.6}$ & $   6.3^{+    4.7}_{-   2.7}$ & $ 630^{+  465}_{-
 267}$ \\ [1ex]
\hline
\end{tabular}
\tablefoot{
\tablefoottext{a}{The SFRs were calculated using the $L_{\rm IR}$ values derived from the DL07 dust model.} 
\tablefoottext{b}{The large uncertainties in the derived values arise from the highly confused \textit{Herschel} photometry for J1000+0234 (see Table~\ref{tab:herschelphot}).} 
}
\end{table*}

\begin{acknowledgements}

We thank the referee for insightful comments on the manuscript. This research was funded by the European Union's Seventh Frame-work program under grant agreement 337595 (ERC Starting Grant, 'CoSMass'). AK acknowledges support by the Collaborative Research Council 956,
sub-project A1, funded by the Deutsche Forschungsgemeinschaft (DFG). The Dark Cosmology Centre is funded by the Danish National Research Foundation. The National Radio Astronomy Observatory is a facility of the National Science Foundation operated under cooperative agreement by Associated Universities, Inc.

\end{acknowledgements}

\end{document}